\documentclass[useAMS]{mn2e}
\usepackage{amssymb}

\usepackage{graphicx}
\title[Asteroid families in the first order resonances]%
{Asteroid families in the first order resonances with Jupiter}
\author[M. Bro\v{z} and D. Vokrouhlick\'y]{M. Bro\v{z}$^{1}$%
 \thanks{E-mail: mira@sirrah.troja.mff.cuni.cz (MB); vokrouhl@cesnet.cz (DV).}
 and D. Vokrouhlick\'y$^{1}$
 \\
$^{1}$Institute of Astronomy, Charles University, Prague, V Hole\v sovi\v ck\'ach 2,
 18000 Prague 8, Czech Republic}

\begin{document}

\date{Accepted ???. Received ???; in original form ???}

\pagerange{\pageref{firstpage}--\pageref{lastpage}} \pubyear{2008}

\maketitle

\label{firstpage}


\begin{abstract}
 Asteroids residing in the first-order mean motion resonances with Jupiter
 hold important information about the processes that set the final
 architecture of giant planets. Here we revise current populations of
 objects in the J2/1 (Hecuba-gap group), J3/2 (Hilda group) and J4/3 (Thule
 group) resonances. The number of multi-opposition asteroids found is
 $274$ for J2/1, $1197$ for J3/2 and $3$ for J4/3. By discovering a second
 and third object in the J4/3 resonance, (186024) 2001~QG$_{207}$ and
 (185290) 2006~UB$_{219}$, this
 population becomes a real group rather than a single object. Using both
 hierarchical clustering technique and colour identification we characterise
 a collisionally-born asteroid family around the largest object (1911)
 Schubart in the J3/2 resonance. There is also a looser cluster around the
 largest asteroid (153) Hilda.
 Using $N$-body numerical simulations we prove that the Yarkovsky effect
 (infrared thermal emission from the surface of asteroids) causes
 a systematic drift in eccentricity for resonant asteroids, while their
 semimajor axis is almost fixed due to the strong coupling with Jupiter.
 This is a different mechanism from main belt families, where the Yarkovsky
 drift affects basically the semimajor axis. We use the eccentricity evolution
 to determine the following ages: $(1.7\pm0.7)\,{\rm Gyr}$ for the Schubart
 family and $\gtrsim 4\,{\rm Gyr}$ for the Hilda family.
 We also find that collisionally-born clusters in the J2/1 resonance
 would efficiently dynamically disperse.
 The steep size distribution of the stable population inside this resonance 
 could thus make sense if most of these bodies are fragments from an event
 older than $\simeq 1$\,Gyr.
 Finally, we test stability of resonant populations during Jupiter's and Saturn's
 crossing of their mutual mean motion resonances. In particular we find
 primordial objects in the J3/2 resonance were efficiently removed
 from their orbits when Jupiter and Saturn crossed their 1:2 mean
 motion resonance.
\end{abstract}




\section{Introduction}
Populations of asteroids in the Jovian first order mean motion resonances 
--J2/1, J3/2 and J4/3-- are closely linked to the orbital evolution of the
giant planets. This is because of their orbital proximity to Jupiter.%
\footnote{Interestingly, at their discovery (158) Hilda and (279) Thule,
 residing in the J3/2 and J4/3 resonances, immediately attracted attention
 of astronomers by vastly extending asteroid zone toward giant planets
 and by their ability to apparently approach Jupiter near aphelia of
 their orbits (e.g., K\"uhnert 1876; Krueger 1889).}
Stability or instability of these asteroid populations directly derives
from the orbital configuration of the giant planets. As such it
is also sensitive on the nature and amount of Jupiter's migration and
other finer details of its dynamics. As a result, the currently
observed asteroids in the Jovian first order resonances contain valuable
information about the early evolution of planets and, if correctly
understood and properly modelled, they may help to constrain it.

Apart from the Trojan clouds (not studied in this paper) the largest 
known population in the Jovian mean motion resonances occupies the 
J3/2 resonance, and is frequently called the Hilda group. It was
carefully studied in a parallel series of papers by Schubart and Dahlgren
and collaborators during the past few decades. Schubart (1982a,b, 1991) 
analysed short-term dynamics of Hilda-type orbits and introduced
quasi-constant orbital parameters that allowed their first classification.
While pioneering, Schubart's work had the disadvantage of having much
smaller sample of known asteroids and computer power than
today. Dahlgren \& Lagerkvist (1995) and Dahlgren et~al. (1997, 1998,
1999) conducted the first systematic spectroscopic and rotation-rate
investigation of Hildas. They found about equal abundance of D- and
P-type asteroids%
\footnote{Note the former P-type objects were reclassified to X-type
 in a newer taxonomy by Bus and Binzel (e.g., Bus et~al. 2002).}
and suggested spectral-size correlation such that
P-types dominate large Hildas and D-type dominate smaller Hildas. They
also suggested small Hildas have large lightcurve amplitudes, as an
indication of elongated or irregular shape, and that the distribution
of their rotation rates is non-Maxwellian. Further analysis using the
Sloan Digital Sky Survey (SDSS) data however does not support significant
dominance of either of the two spectral types for small sizes and
indicates about equal mix of them (Gil-Hutton \& Brunini 2008; see also
below). Smaller populations of asteroids in the J2/1 and J4/3 received
comparatively less observational effort.

Since the late 1990s powerful-enough computers allowed a more
systematic analysis of fine details of the longer-term dynamics
in the Jovian first order resonances. Ferraz-Mello \&
Michtchenko (1996) and Ferraz-Mello et~al. (1998a,b) determined that
asteroids in the J2/1 resonance can be very long-lived, possibly
primordial, yet their motion is comparatively more chaotic than those
in the J3/2 resonance. The latter paper showed that commensurability
between the libration period and the period of Jupiter's and Saturn's 
Great Inequality might have played an important role in
depletion of the J2/1 resonance. This would have occurred when both
giant planets were farther from their mutual 2:5 mean motion
configuration in the past. A still more complete analysis was
obtained by Nesvorn\'y \& Ferraz-Mello (1997) who also pointed out
that the J4/3 resonance stable zone is surprisingly void of asteroids,
containing only (279) Thule. Roig et~al. (2001) and Bro\v{z} et~al.
(2005) recently revised the population of asteroids in the J2/1 resonance
and classified them into several groups according to their long-term
orbital stability. While the origin of the unstable resonant population was
successfully interpreted using a model of a steady-state flow of main belt
objects driven by the Yarkovsky semimajor axis drift, the origin of the
long-lived asteroids in the J2/1 remains elusive. Population of Hildas and Thule
was assumed primordial or captured by an adiabatic migration of Jupiter
(e.g., Franklin et~al. 2004).

It has been known for some time that the current configuration of
giant planets does not correspond to that at their birth. However,
a new momentum to that hypothesis was given by the so called Nice
model (Tsiganis et~al. 2005; Morbidelli et~al. 2005; Gomes et~al.
2005). The Nice model postulates the initial configuration of the
giant planets was such that Jupiter and Saturn were interior of their
mutual 1:2 mean motion resonance (see also Morbidelli et~al. 2007).
The event of crossing this resonance had a major influence on the final
architecture of giant planets and strongly influenced structure
of small-bodies populations in the Solar system. Morbidelli et~al.
(2005) showed that the population of Jupiters Trojan asteroids was
destabilised and re-populated during this phase. In what follows we
show that, within the Nice model, the same most probably occurs
for populations  of asteroids in the J3/2 and J4/3 resonances.

The paper is organised as follows: in Section~\ref{popul} we revise information 
about the current populations of asteroids in the Jovian first order 
resonances. We use an up-to-date {\tt AstOrb} database of asteroid 
orbits from the Lowell Observatory ({\tt ftp.lowell.edu}) as of September~2007
and eliminate only single-opposition cases to assure accurate orbital information.

In Section~\ref{fams} we apply clustering techniques and extract two families
of asteroids on similar orbits in the J3/2 resonance. We strengthen their
case with an additional colour analysis using the SDSS broadband data.
We model the long-term orbital evolution of these families and estimate
their ages on the basis of Yarkovsky-driven dispersion in eccentricity.

In Section~\ref{stability} we determine an orbital stability of the putative 
primordial populations of planetesimals in the Jovian first-order resonances.
We show that those in the J3/2 and J4/3 are very efficiently eliminated when
Jupiter and Saturn cross their mutual 1:2 mean motion resonance.
We also determine the removal rate of very small resonant asteroids
due to the Yarkovsky/YORP effects.


\section{Current asteroid populations in the Jovian first order
 resonances}\label{popul}

Dynamics of asteroid motion in the Jovian first order resonances
has been extensively studied by both analytical and numerical methods 
in the past few decades (e.g., Murray 1986; Sessin \& Bressane 1988; 
Ferraz-Mello 1988; Lemaitre \& Henrard 1990; Morbidelli 
\& Moons 1993; Moons et~al. 1998; Nesvorn\'y \& Ferraz-Mello 1997;
Roig et~al. 2002; Schubart 2007). In what follows we review a minimum
information needed to understand our paper, referring an interested 
reader to the literature mentioned above for more insights.

In the simplest framework of a circular restricted planar three-body
problem (Sun-Jupiter-asteroid) the fundamental effects of the resonant 
dynamics is reduced to a one-degree of freedom problem defined
by a pair of variables $(\Sigma,\sigma)$. For J$(p+1)/p$ resonance 
($p=1,2$ and $3$ in our cases) we have
\begin{eqnarray}
 \Sigma &=& \sqrt{a}\,\left(1-\sqrt{1-e^2}\right)\; , \label{res11} \\
 \sigma &=& \left(p+1\right)\lambda' - p\, \lambda - \varpi \; ,
 \label{res12}
\end{eqnarray}
where $a$ is the semimajor axis, $e$ the eccentricity, $\varpi$ the 
longitude of pericentre and $\lambda$ the mean longitude in orbit of 
the asteroid, and $\lambda'$ is the mean longitude in orbit of Jupiter.

If the asteroid motion is not confined into the orbital plane of the
planet, we have an additional pair of resonant variables $(\Sigma_z,
\sigma_z)$ such that
\begin{eqnarray}
 \Sigma_z &=& 2\sqrt{a\left(1-e^2\right)}\sin^2 \frac{i}{2}\; ,
  \label{res21} \\
 \sigma_z &=& \left(p+1\right)\lambda' - p\, \lambda - \Omega \; ,
  \label{res22}
\end{eqnarray}
where $i$ denotes the inclination of asteroids orbit and $\Omega$ the longitude
of its node. Remaining still with the simple averaged model,
orbital effects with shorter periods are neglected, the motion obeys
an integral of motion $N$ given by
\begin{equation}
 N=\sqrt{a}\,\left(\frac{p+1}{p}-\sqrt{1-e^2}\cos i\right)\; .
  \label{resint}
\end{equation}
Because of this integral of motion, variations
of $\Sigma$ imply oscillations of both $a$ and $e$. 

The two-degree of freedom character of the resonant motion
prevents integrability. However, as an approximation we may introduce
a hierarchy by noting that perturbation described by the $(\Sigma,
\sigma)$ variables is larger than that described by the $(\Sigma_z,
\sigma_z)$ terms (e.g., Moons et~al. 1998). This is usually true for
real resonant asteroids of interest. Only the angle $\sigma$ librates
and $\sigma_z$ circulates with a very long period. The $(\Sigma_z, 
\sigma_z)$ dynamics thus produces a long-period perturbation of 
the $(\Sigma,\sigma)$ motion.

Within this model the minimum value of $\Sigma$ in one resonant cycle
(typically several hundreds of years) implies $a$ is minimum and 
$e$ is maximum. These values do not conserve exactly from one cycle 
to another because the $(\Sigma_z, \sigma_z)$ motion produces small
oscillations. Since $\Sigma+\Sigma_z-N=-\sqrt{a}/p$ one needs
to wait until $\Sigma_z$ reaches maximum over its cycle to attain
`real' minimum of $a$ values and `real' maximum of $e$ values over a 
longer time interval. From (\ref{res21}) we note the maximum of $\Sigma_z$ 
occurs for the maximum of $i$ variations. This situations occurs typically
once in a few thousands of years. In an ideal situation, these extremal
values of $(a,e,i)$ would be constant and may serve as a set of proper
orbital elements.

The motion of real asteroids in the Solar system is further
complicated by Jupiter having non-zero and oscillating value of
eccentricity. This brings further perturbations (e.g., Ferraz-Mello
1988; Sessin \& Bressane 1988 for a simple analytic description)
and sources of instability inside the resonance. Despite the
non-integrability, we follow Roig et~al. (2001) and introduce
{\em pseudo-proper orbital elements\/} $(a_p,e_p,\sin i_p)$
as the osculating elements $(a,e,\sin i)$ at the moment, when the
orbit satisfies the condition:
\begin{equation}
 \sigma=0 \land {d\sigma\over dt}<0 \land \varpi-\varpi'=0 \land
 \Omega-\Omega'=0\; , \label{prop1}
\end{equation}
where $\varpi'$ and $\Omega'$ denote the longitude of pericentre
and the longitude of node
of Jupiter. As above, when (\ref{prop1}) holds the osculating
orbital elements are such that $a$ attains minimum, $e$
attains maximum and $i$ attains maximum. Numerical experiments
show that with a complete perturbation model and a finite time-step
it is difficult to satisfy all conditions of (\ref{prop1})
simultaneously. Following Roig et~al. (2001) we thus relax
(\ref{prop1}) to a more practical condition
\begin{equation}
 |\sigma|<5^\circ \land {\Delta\sigma\over \Delta t}<0 \land
 |\varpi-\varpi'|<5^\circ\; . \label{prop2}
\end{equation}
Because this condition is only approximate, we numerically integrate
orbits of resonant asteroids for 1\,Myr, over which the pseudo-proper
orbital elements are recorded. We then compute their mean value
and standard deviation, which is an expression of the orbital
stability over that interval of time.

In the case of the J3/2 and J4/3 resonances, we use the condition
(\ref{prop2}) with a different sign ${\Delta\sigma\over \Delta t}>0$ 
and, moreover, we apply a digital filter (denoted as~A in Quinn et~al.
(1991), using 1\,yr sampling and a decimation factor of~10) to 
$\sigma(t)$. This intermediate stage serves to suppress oscillations
faster than the libration period. The different sign of ${\Delta\sigma\over
\Delta t}$ just means our pseudo-proper orbital elements correspond
to maximum value of $a$ and minimum values of $e$ and $i$, in order
to allow more direct comparison with previous analyses.

Aside to this short-term integration we perform long-term
runs to determine the stability of a particular resonant orbit.
With this aim we conduct integrations spanning 4\,Gyr for all resonant
asteroids. Because of the inherent uncertainty in the initial conditions
(orbital elements at the current epoch), we perform such integration
for the nominal orbit and $10$ clones that randomly span the uncertainty
ellipsoid. We then define {\em dynamical lifetime\/} of the orbit
as the median of time intervals, for which the individual
clones stayed in the resonance.

All integrations are performed using the {\sc Swift} package 
(Levison \& Duncan 1994), slightly modified to include necessary
on-line digital filters and a second order symplectic integrator
(Laskar \& Robutel 2001). Most of numerical simulations take into
account gravitational interactions only, but in specific cases 
-- and when explicitly mentioned -- we include also Yarkovsky (thermal)
accelerations. In this case we use an implementation described
in detail by Bro\v z (2006). Our simulations include 4 outer planets.
We modify the initial conditions of the planets and asteroids by a barycentric 
correction to partially account for the influence of the terrestrial planets.
The absence of the terrestrial planets as perturbers is a reasonable
approximation in the outer part of the main belt and for orbits with
$e<0.8$ in general. We nevertheless checked the short-term computations
(determination of pseudo-proper resonant elements) using a complete
planetary model and noticed no significant difference in results. The
second order symplectic scheme allows us to use a time step of 91~days.


\subsection{Hecuba-gap group}\label{hecuba}

In order to determine, which objects are located in the J2/1 mean motion resonance,
we first extracted orbits from the {\tt AstOrb} database with osculating orbital
elements in a broad box around this resonance (see, e.g., Roig et~al.
2001 for a similar procedure). We obtained 7139 orbits, which
we numerically integrated for 10\,kyr. We recorded and analysed
behaviour of the resonance angle $\sigma=2\lambda'-\lambda-\varpi$ from
Eq.~(\ref{res12}). Pericentric librators, for which $\sigma$ oscillates
about $0^\circ$, were searched. We found 274 such cases; this extends
the previous catalogue of Bro\v{z} et~al. (2005) almost twice. The
newly identified resonant objects are mainly asteroids discovered or 
recovered after 2005 with accurate enough orbits. We disregard from our 
analysis asteroids at the border of the resonance, for which $\sigma(t)$ 
exhibits alternating periods of libration and circulation, and also those
asteroids for which $\sigma$ oscillates but are not resonant anyway
($N\leq 0.8$ in Eq.~(\ref{resint}); see, e.g., Morbidelli \& Moons 1993).
The latter reside on low-eccentricity orbits in the main
asteroid belt adjacent to the J2/1 resonance.

\begin{figure}
 \includegraphics[width=84mm]{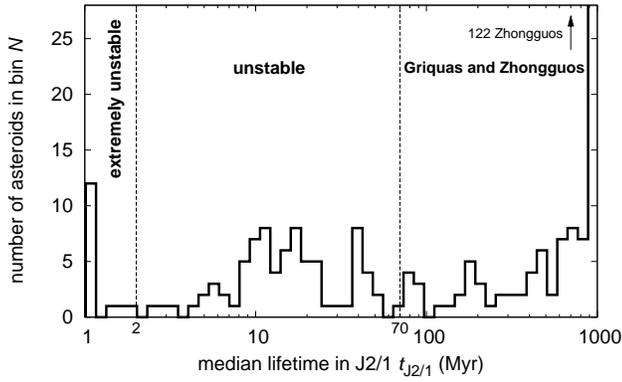}
\caption{A distribution of the median dynamical lifetimes for objects
  in the J2/1 resonance. A division to several groups is denoted:
  extremely unstable objects ($t_{\rm J2/1} \le 2\,{\rm Myr}$),
  short-lived objects ($t_{\rm J2/1} \le 70\,{\rm Myr}$)
  and long-lived objects (Griquas and Zhongguos, $t_{\rm J2/1} > 70\,
  {\rm Myr}$).}
\label{r21_obs_loglftime4}
\end{figure}

We conducted short- and long-term integrations of the resonant asteroids
as described above. They allowed us to divide the population
into 182~long-lived asteroids (with the median dynamical lifetime
longer than $70$\,Myr, as defined in Bro\v z et~al. 2005) and 92~short-lived asteroids
(the lifetime shorter than $70$\,Myr), see Fig.~\ref{r21_obs_loglftime4}.%
\footnote{Our results for both J2/1 and J3/2 resonances are summarised
 in tables available through a website {\tt http://sirrah.troja.mff.cuni.cz/yarko-site/} (those for J4/3 bodies
 are given in Table~\ref{tab1}). These contain
 listing of all resonant asteroids, their pseudo-proper orbital
 elements with standard deviations, their dynamical residence time
 and some additional information.}
Among the short-lived objects we found 14 have
dynamical lifetimes even less than $2$\,Myr and we call them extremely unstable.
Bro\v{z} et~al. (2005) suggested the unstable orbits in the J2/1
resonance are resupplied from the adjacent main belt due to a
permanent flux driven by the Yarkovsky force, the extremely 
unstable objects are most probably temporarily captured Jupiter-family comets.
The origin of the long-lived population in this resonance is still not known.

\begin{figure}
\begin{minipage}{84mm}
 \includegraphics[width=84mm]{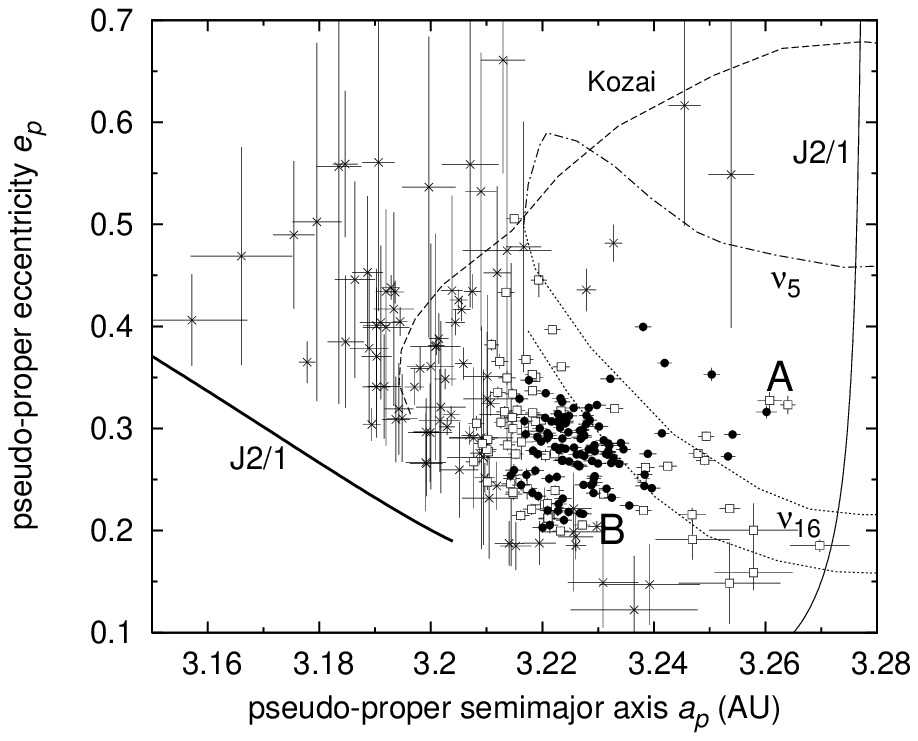}\\
 \includegraphics[width=84mm]{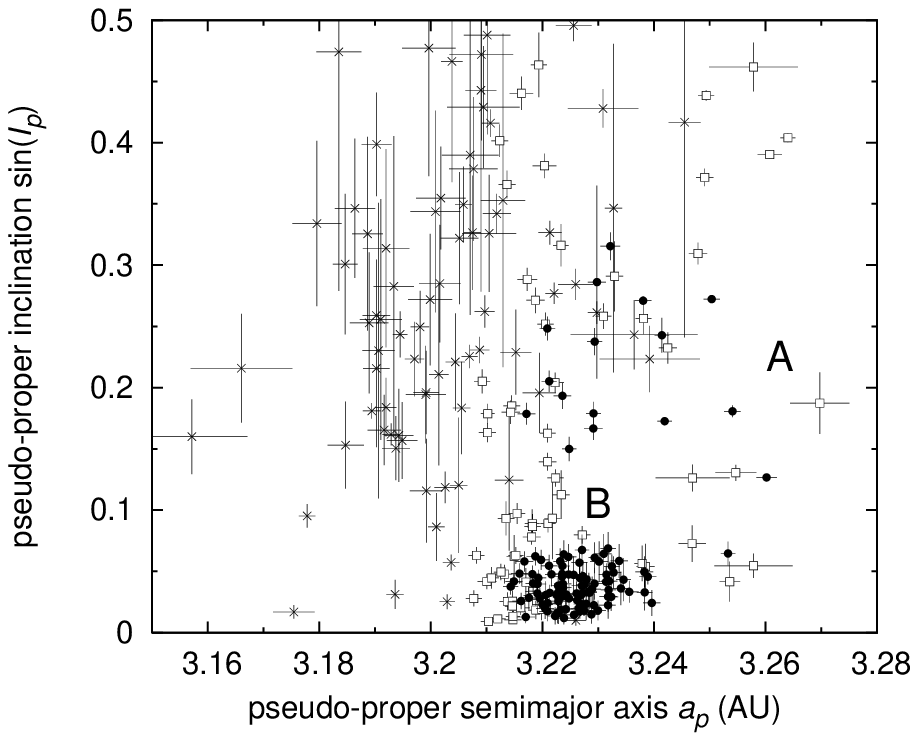}
\end{minipage}
 \caption{Pseudo-proper orbital elements for the 247 objects in the
   J2/1 resonance projected onto the planes of semimajor axis
   $a_p$ vs eccentricity $e_p$ (top) and semimajor axis $a_p$ vs
   sine of inclination $\sin i_p$ (bottom). Bars are standard
   deviations of the elements derived from 1~Myr numerical
   integration. Position of several secular resonances embedded
   in J2/1 is shown in the upper panel. The unstable population
   of asteroids (crosses) occupies the region of their overlap; the
   stable population (full circles) occupies two distinct zones
   --A and B-- of low-eccentricity and low-inclination orbits 
   (e.g., Nesvorn\'y \& Ferraz-Mello 1997). The population of
   marginally stable asteroids (open squares) resides in region
   adjacent to the unstable borders of the resonance or near the
   bridge over the stable regions associated with the $\nu_{16}$
   secular resonance.}
 \label{r21proper}
\end{figure}

Figure~\ref{r21proper} shows the pseudo-proper orbital elements of the J2/1 
asteroids projected onto the $(a_p,e_p)$ and $(a_p,\sin i_p)$ planes. 
Our data confirm that the unstable population of J2/1 asteroids
populates the resonance outskirts near its separatrix, where
several secular resonances overlap and trigger chaotic motion
(e.g., Morbidelli \& Moons 1993; Nesvorn\'y \& Ferraz-Mello 1997;
Moons et~al. 1998). At low-eccentricities the chaos is also caused
by an overlap on numerous secondary resonances (e.g., Lemaitre \&
Henrard 1990). Two `islands' of stability --A and B-- harbour the
long-lived population of bodies. The high-inclination island A, separated
{}from the low-inclination island B by the $\nu_{16}$ secular resonance,
is much less populated. Our current search identifies 9~asteroids in
the island A. The origin of the asymmetry in A/B islands is not known, but since
the work of Michtchenko \& Ferraz-Mello (1997) and Ferraz-Mello et~al.
(1998a,b) it is suspected to be caused by instability due to the libration 
period commensurability with the forcing terms produced by the Great
Inequality.

\begin{figure}
\begin{center}
 \includegraphics[width=74mm]{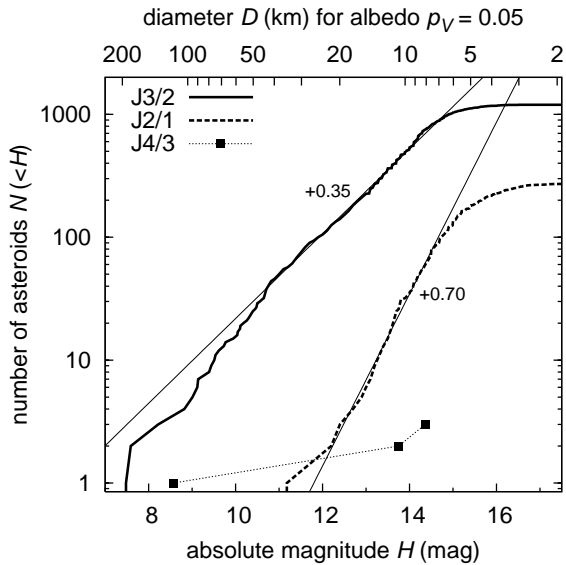}
\end{center}
\caption{Cumulative distributions $N({<}H)$ of the absolute magnitude
  $H$ values for population of asteroids in the Jovian first order mean
  motion resonances: (i) J2/1 (dashed curve), (ii) J3/2 (solid curve),
  and (iii) J4/3 (dotted curve). The straight lines show best-fit
  approximations $N({<}H)\propto 10^{\gamma H}$ with the values of
  $\gamma$ indicated by the corresponding label. The fit matches $N({<}H)$
  for $H$ in the interval $(12, 15)\,{\rm mag}$ for J2/1 and $(10.5, 14.5)
  \,{\rm mag}$ for J3/2;
  no such approximation is available for J4/3 where only three objects
  are currently known. The $H$ values where the straight line
  approximations level off from the data roughly correspond to the
  completeness limit of the population (R.~Jedicke, personal
  communication). For sake of a rough comparison, the upper abscissa
  gives an estimate of sizes for the albedo value $p_V=0.05$, average
  of the outer belt population.}
 \label{ressfd}
\end{figure}

The size-frequency distribution of objects of a population
is an important property, complementing that of the orbital
distribution. Figure~\ref{ressfd} shows cumulative distribution 
$N({<}H)$ of the absolute magnitudes $H$ for bodies in the J2/1
(and other Jovian first order resonances as well). 
In between $H=12\,{\rm mag}$ and $14.5\,{\rm mag}$ (an approximate
completeness limit;
R.~Jedicke, personal communication) it can be matched by a simple
power-law $N({<}H)\propto 10^{\gamma H}$, with $\gamma = (+0.70\pm0.02)$.%
\footnote{This is equivalent to a cumulative size distribution law
 $N({>}D)\propto D^\alpha$ with $\alpha=-5\gamma = (-3.5\pm0.1)$,
 assuming all bodies have the same albedo.}
We thus confirm that the J2/1 population is steeper than it would
correspond to a standard collisionally evolved system (e.g.,
Dohnanyi 1969; O'Brien \& Greenberg 2003) with $\gamma=+0.5$.
The same result holds for both the short- and long-lived sub-populations
in this resonance separately.

\begin{figure}
 \includegraphics[width=84mm]{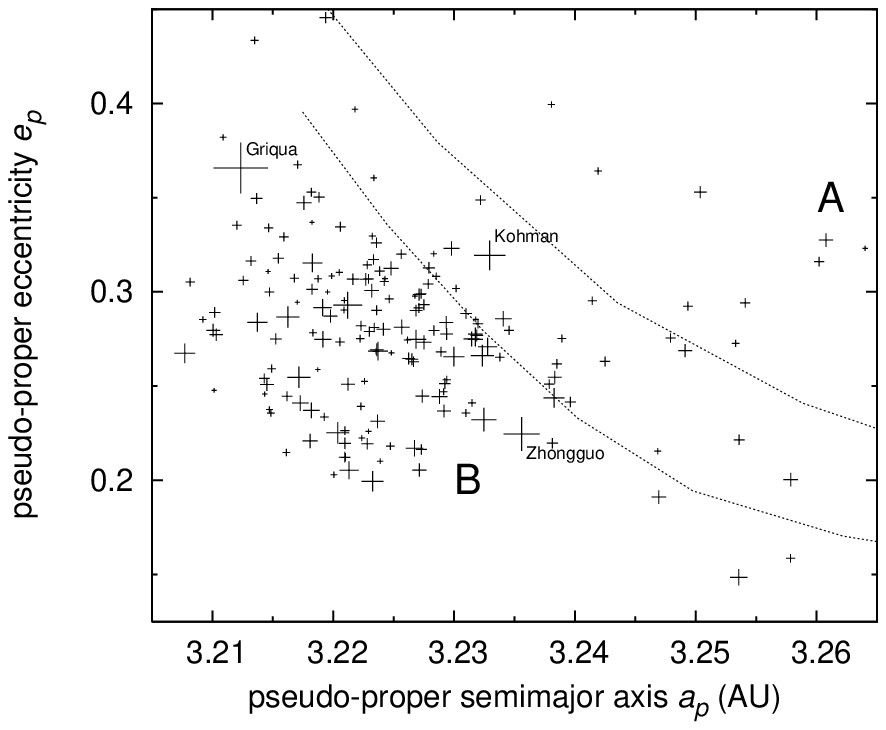}\\
 \includegraphics[width=84mm]{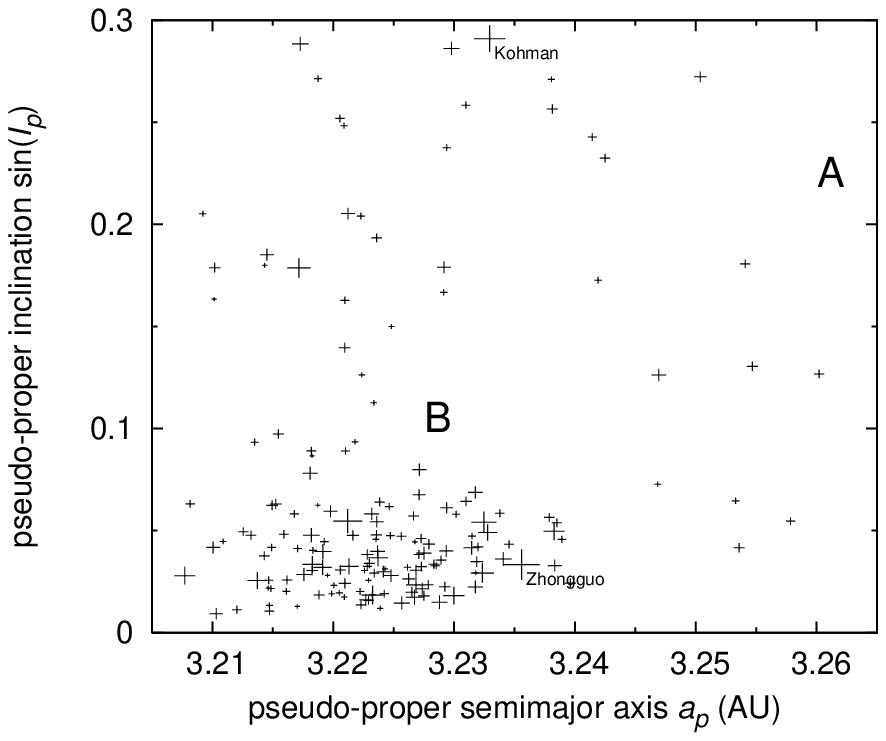}
\caption{A zoom on the $(a_p,e_p)$ and $(a_p,\sin i_p)$ plots from
  Fig.~\ref{r21proper} with relative size of the resonant asteroids
  indicated by size of the crosses. Note the large bodies, some of which
  are labelled, reside far from each other.}
 \label{r21_arerir_sizes}
\end{figure}

Albedos of J2/1 bodies are not known, except for (1362)~Griqua for which
Tedesco et~al. (2002) give $p_V = 0.067$. The surroundings main-belt population 
has an average $p_V = 0.05$. For sake of simplicity we convert absolute 
magnitudes to sizes using this averaged value when needed. For instance
in Fig.~\ref{r21_arerir_sizes} we show a zoom on the long-lived population
of objects in the J2/1 resonance with symbol size weighted by the 
estimated size of the body. We note large objects are located far from 
each other and they are quite isolated --- no small asteroids are in
close surroundings. Both these observations suggest that
the long-lived J2/1 population does not contain recently-born collisional
clusters.

\begin{figure}
\begin{minipage}{84mm}
 \includegraphics[width=84mm]{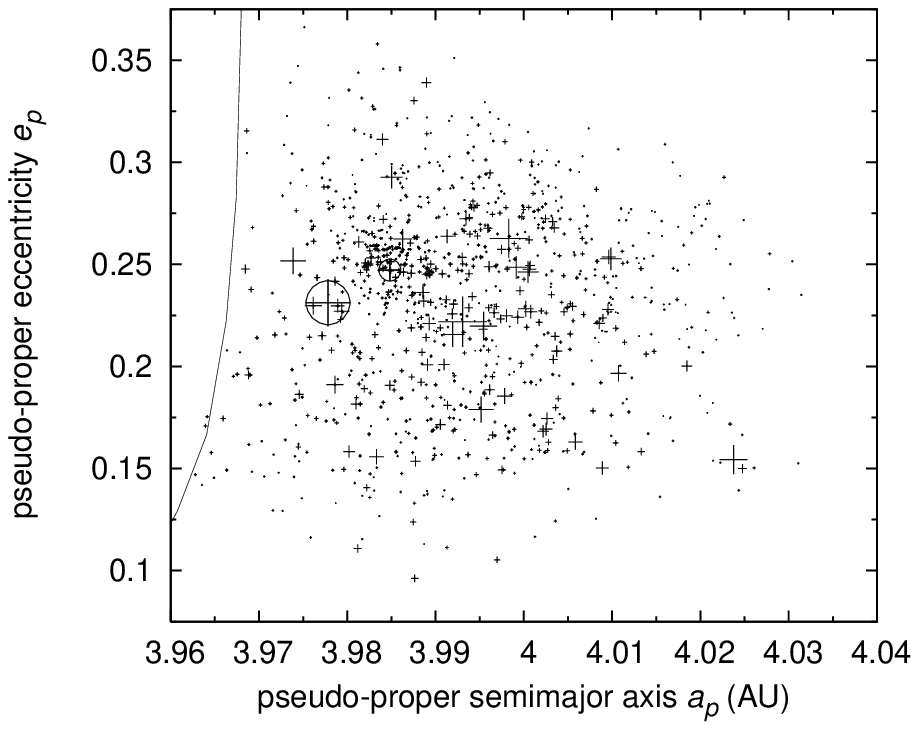}\\
 \includegraphics[width=84mm]{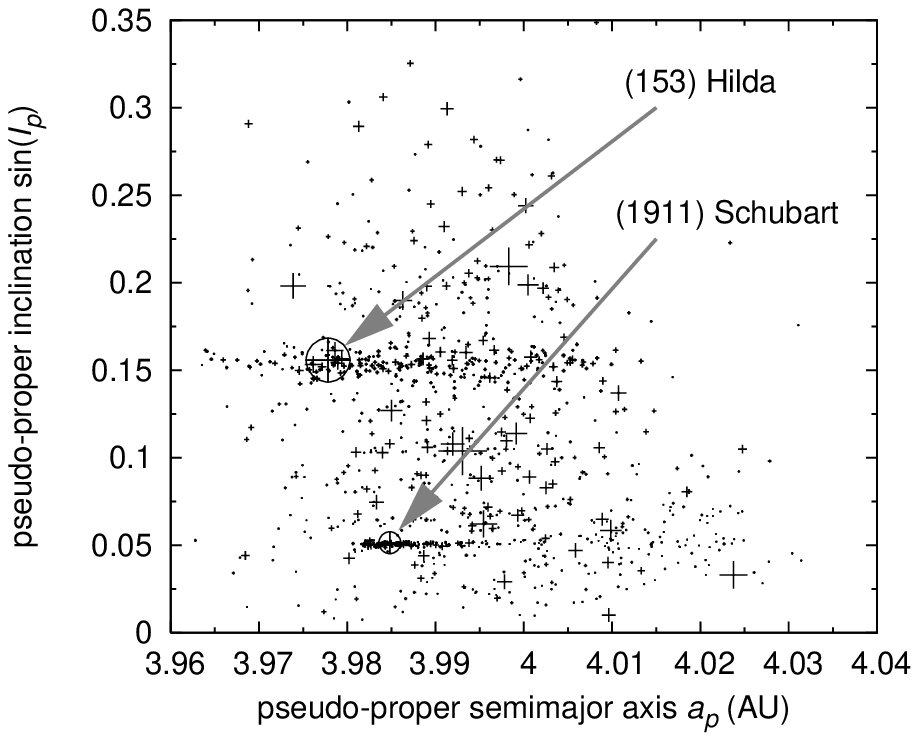}
\end{minipage}
 \caption{Pseudo-proper orbital elements for 1197 Hildas projected onto
  the planes of semimajor axis $a_p$ vs eccentricity $e_p$ (top) and
  semimajor axis $a_p$ vs sine of inclination $\sin i_p$ (bottom).
  Larger size of the symbol indicates larger physical size of the
  asteroid. Because of Hildas orbital stability, the uncertainty in the
  pseudo-proper element values is typically smaller than the symbol
  size. Note a tight cluster around the proper inclination value
  $\sin i_p\simeq 0.0505$, lead by the largest asteroid (1911) Schubart,
  and a somewhat looser cluster around the proper inclination value
  $\sin i_p\simeq 0.151$, lead by the largest asteroid (153) Hilda.
  Both are discussed in more detail in Section~\ref{fams}.
  Solid line denotes the libration centre of the J3/2 resonance.}
 \label{r32proper}
\end{figure}

\begin{figure}
 \includegraphics[width=84mm]{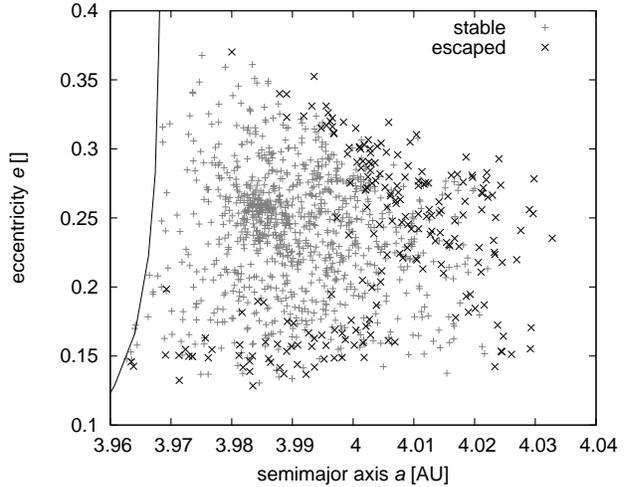}
\caption{Pseudo-proper semimajor axis vs eccentricity plot for asteroids
  in the J3/2 resonance. Bold crosses denote orbits, which escaped during
  the 4\,Gyr of evolution. Solid line is the position of the libration
  centre (the outer separatrix is located further to the right).}
 \label{r32_ae_escape}
\end{figure}

\begin{figure}
\begin{center}
 \includegraphics[width=74mm]{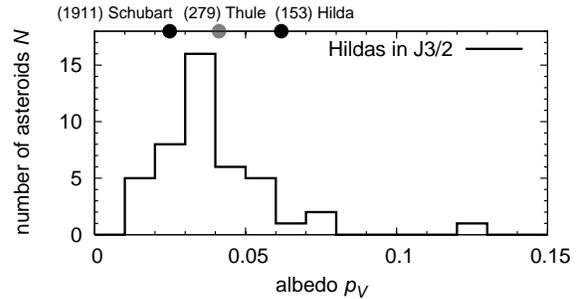}
\end{center}
\caption{The histogram shows distribution of values of the geometric
  albedo $p_V$ determined by Tedesco et~al. (2002) for asteroids located
  inside the J3/2 resonance. Three individual values -- (153)~Hilda,
  (1911)~Schubart and (279)~Thule (a J4/3 asteroid) -- are indicated on
  top.}
 \label{resalb}
\end{figure}


\subsection{Hilda group}

Because asteroids in the J3/2 constitute a rather isolated group,
it is easy to select their candidates: we simply extracted from the
{\tt AstOrb} database those asteroids with semimajor axis in
between $3.8$~AU and $4.1$~AU. With that we obtained 1267 multi-opposition
objects. We numerically integrated these orbits for 10\,kyr and analysed
the behaviour of the resonance angle $\sigma=3\lambda'-2\lambda-\varpi$.
We obtained 1197 cases for which $\sigma$ librates about $0^\circ$
and which have $N\geq 0.44$, a threshold of the resonance zone
(e.g., Morbidelli \& Moons 1993); see Fig.~\ref{r32proper}.

The long-term evolution of Hildas indicates that not all of them
are stable over 4\,Gyr, but 20\,\% escape earlier. A brief inspection
of Fig.~\ref{r32_ae_escape} shows, that the escapees are essentially
asteroids located closer to the outer separatrix and exhibiting large
amplitudes of librations. If the Hilda group has been
constituted during the planetary formation some 4 Gyr ago,
some non-conservative process must have placed these objects onto
their currently unstable orbits. We suspect mutual collisions or
gravitational scattering on the largest Hilda members might be the
corresponding diffusive mechanisms. Small enough members might be also
susceptible to the resonant Yarkovsky effect (see Sec.~\ref{yarkovsky}
and the Appendix~A).

Data in Fig.~\ref{ressfd} confirm earlier findings that the Hilda 
group is characterized by an anomalously shallow size distribution.
In between absolute magnitudes $H=10.5\,{\rm mag}$ and $14.5\,{\rm mag}$ 
the cumulative distribution can be well matched by $N({<}H)\propto 
10^{\gamma H}$ with $\gamma = (+0.35\pm0.02)$ only.

Sub-populations among Hilda asteroids, namely two collisional families,
are studied in Section~\ref{fams}.


\subsection{Thule group}

In spite of a frequent terminology ``Thule group'', asteroids in the
J4/3 resonance consisted of a single object (279) Thule up to now.
Nesvorn\'y \& Ferraz-Mello (1997) considered this situation anomalous
because the extent of the stable zone of this resonance is not much
smaller than that of the J3/2 resonance (see also Franklin et~al.
2004). In the same way, our knowledge about the low-$e$ and low-$i$ 
Thule-type stable orbits (e.g., $a=4.27$\,AU, $e=0.1$ and $i=5^\circ$) 
should be observationally complete at about magnitudes $H=12.5\mbox{--}13\,{\rm mag}$ 
(R.~Jedicke, personal communication). A rough estimate also shows that
even one magnitude in $H$ beyond this completeness limit the Thule
population should be known at $\sim 10$\% completeness, leaving only 
about $90$\,\% undiscovered population. We thus conclude that the objects
in the magnitude range $H=9\mbox{--}13\,{\rm mag}$ are very likely missing in this resonance.
Where does the existing population of small Thule-type asteroids
begin?

Our initial search in the broad box around the J4/3 resonance detected only 13 
objects. Six of them, including the well-known extinct comet (3552) Don
Quixote (e.g., Weissman et~al. 2002), are on typical orbits of Jupiter-family
comets that happen to reside near this resonance with very
high eccentricity and moderately high inclination. Two more are
single-opposition objects and one has only poorly constrained orbit,
leaving us with (279) Thule and three additional candidate objects:
(52007) 2002~EQ47, (186024) 2001~QG$_{207}$ and (185290) 2006~UB$_{219}$.

\begin{table*}
 \centering
 \begin{minipage}{140mm}
  \caption{Data on presently known population of asteroids residing in
   the J4/3 Jovian mean motion resonance (Thule group). Pseudo-proper
   orbital elements $(a_p,e_p,\sin i_p)$ are given together with their
   standard deviations $(\delta a_p,\delta e_p,\delta \sin i_p)$ 
   determined from a $1$~Myr numerical integration. $\sigma_{p,\,max}$ is the
   maximum libration amplitude in the Sessin's $(K,H)$ variables (see
   Fig.~\ref{43reson}), $H$ is the absolute magnitude from
   the {\tt AstOrb} catalogue and $D$ is the estimated size using $p_V=0.04$
   geometric albedo (Tedesco et~al. 2002). \label{tab1}}
  \begin{tabular}{rlcccccccrr}
  \hline
   No. & Name & $a_p$ & $e_p$ & $\sin i_p$ & $\delta a_p$ &
   $\delta e_p$ & $\delta \sin i_p$ & $\sigma_{p,\,max}$ & 
   \multicolumn{1}{c}{$H$} & \multicolumn{1}{c}{$D$} \\ [3pt]
   & & [AU] & & & [AU] & & & [deg] & \multicolumn{1}{c}{[mag]} &
   \multicolumn{1}{c}{[km]} \\
 \hline
  279  & Thule           & 4.2855 & 0.119 & 0.024 & 0.0005 & 0.012 &
  0.003 & ${\sim}50$       & $ 8.57$ & $126.6$ \\
186024 & 2001~QG$_{207}$ & 4.2965 & 0.244 & 0.042 & 0.0003 & 0.014 &
  0.003 & $\phantom{\sim}$25 & $14.36$ & $  8.9$ \\
185290 & 2006~UB$_{219}$ & 4.2979 & 0.234 & 0.102 & 0.0003 & 0.014 &
 0.004 & $\phantom{\sim}$25  & $13.75$ & $ 11.8$ \\
\hline
\end{tabular}
\end{minipage}
\end{table*}


\begin{figure*}
\begin{minipage}{176mm}
\begin{center}
 \begin{tabular}{ccc}
  \includegraphics[width=44mm]{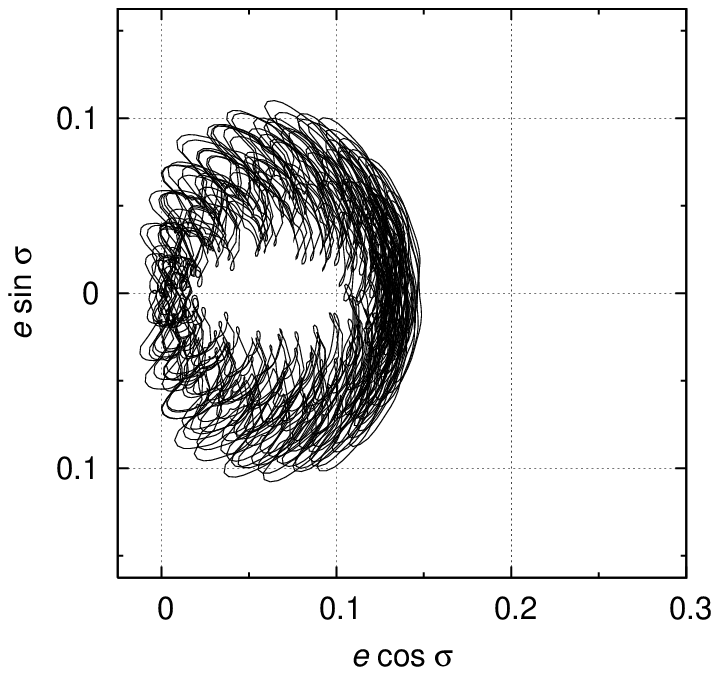} &
  \includegraphics[width=44mm]{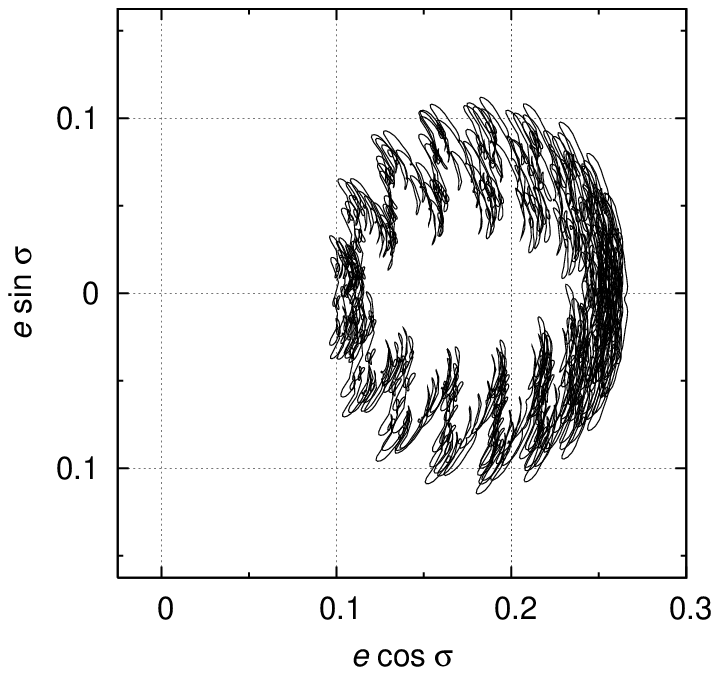} &
  \includegraphics[width=44mm]{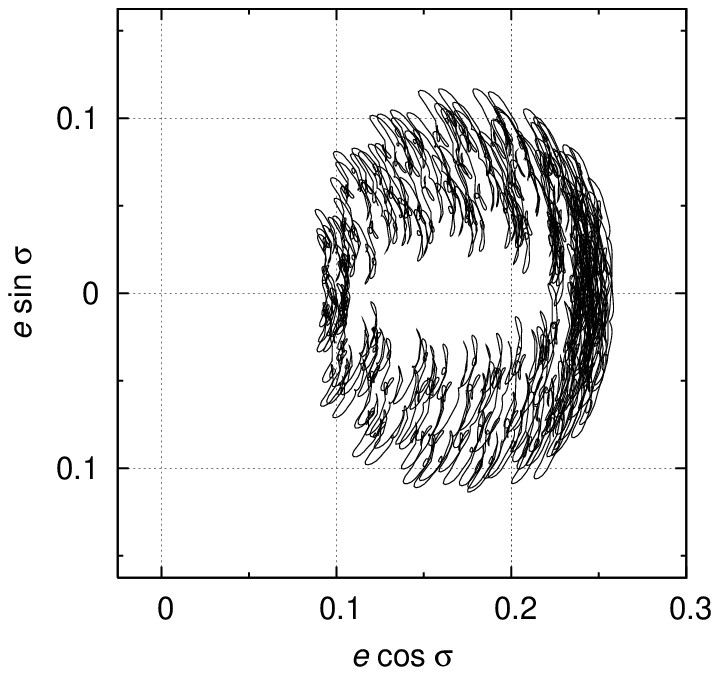} \\
  \includegraphics[width=44mm]{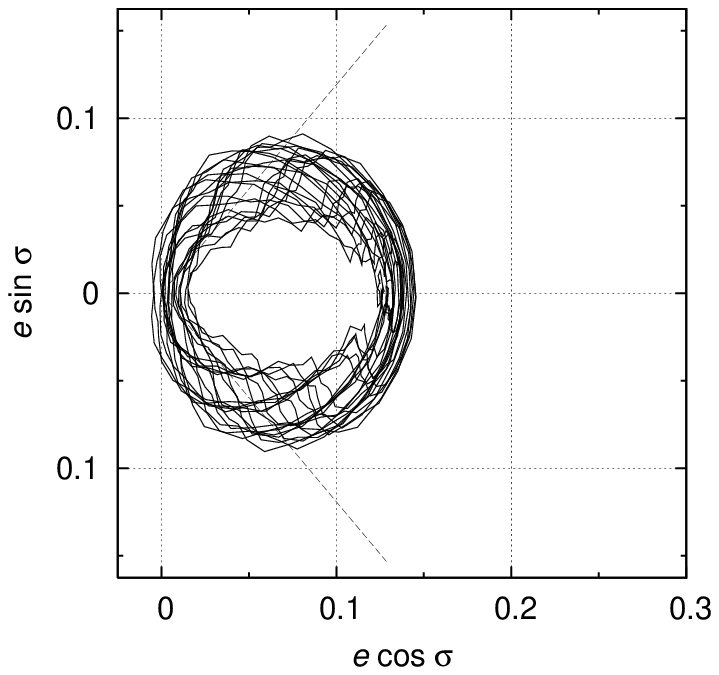} &
  \includegraphics[width=44mm]{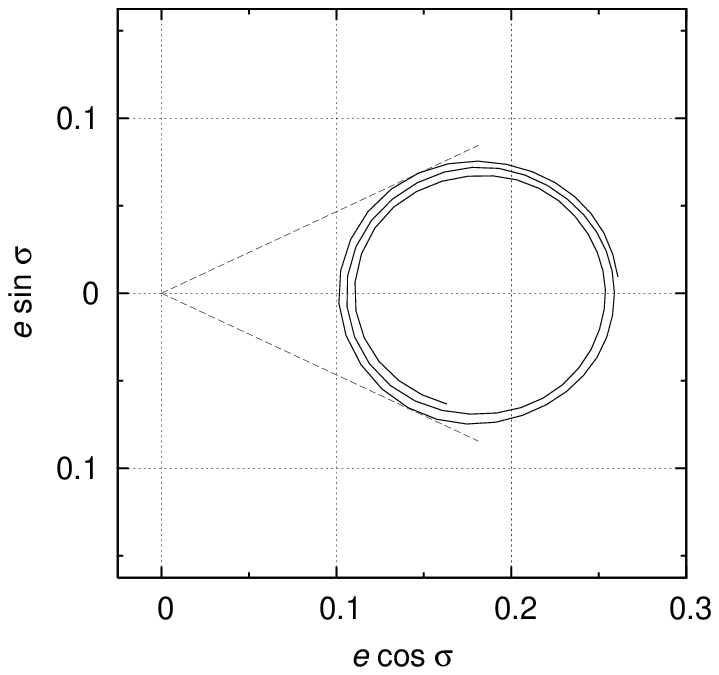} &
  \includegraphics[width=44mm]{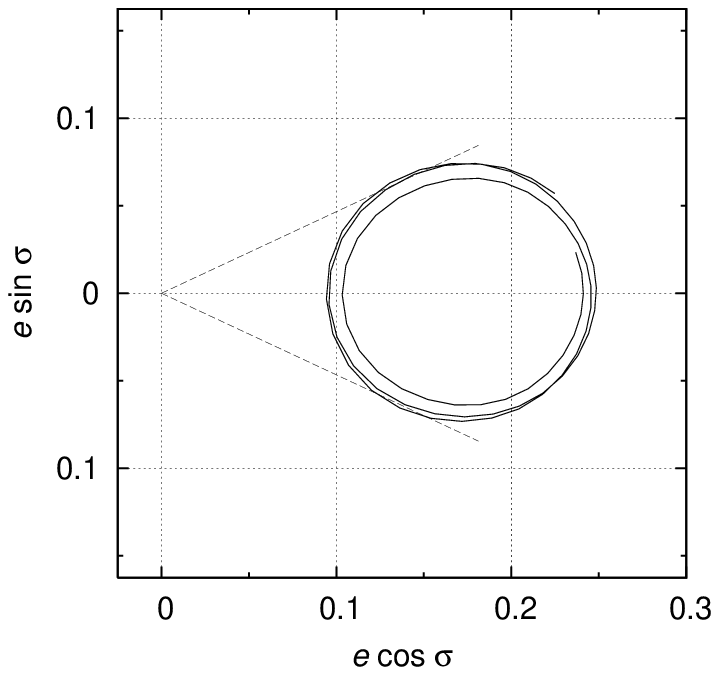} \\
 \end{tabular}
\end{center}
\end{minipage}
\caption{Top panels: orbits of (279)~Thule [left], (186024) 2001~QG$_{207}$
  [middle] and (185290) 2006~UB$_{219}$ [right] in resonant variables
  $(e\cos\sigma,e\sin\sigma)$ of the J4/3 resonance. Here $e$ is the
  eccentricity and $\sigma=4\lambda'- 3\lambda-\varpi$, with $\lambda$ and
  $\lambda'$ the mean longitude in orbit of the asteroid and Jupiter and
  $\varpi$ is the longitude of asteroid's pericentre (in all cases the
  osculating orbital elements are used). Each panel shows results of a short-term,
  10~kyr numerical integration. In all three cases the orbits librate about
  the pericentric branch ($\sigma=0^\circ$) of the resonance. Perturbations
  due to Jupiter's eccentricity and its variations make the regular libration
  move in an epicyclic manner (e.g., Ferraz-Mello 1988; Sessin \& Brassane 1988).
  Bottom panels: filtered resonant variables $(e\cos\sigma, e\sin\sigma)$,
  with short-period oscillations forced by Jupiter removed by digital filtering.
  They are similar to Sessin's $(K,H)$ coordinates in Sessin \& Brassane (1988)
  or Tsuchida (1990). In each of the cases the pericentric libration is clearly
  revealed. The maximum libration amplitude in these coordinates is denoted by
  $\sigma_{p,\,max}$ in Table~\ref{tab1}.}
 \label{43reson}
\end{figure*}

Figure~\ref{43reson} (top panels) shows short-term tracks of (279) Thule, 
(186024) 2001~QG$_{207}$ and (185290) 2006~UB$_{219}$ in resonant variables
$\sqrt{2\Sigma}\,(\cos\sigma,\sin\sigma) \simeq (e\cos\sigma,e\sin\sigma)$
of the J4/3 resonance ($\sigma=4\lambda'-3\lambda-\varpi$ in this case). 
In all cases their orbits librate about the pericentric 
branch ($\sigma=0^\circ$) of this resonance, although this is complicated --mainly
in the low-eccentricity case of (279) Thule-- by the forced terms due 
to Jupiter's eccentricity (see, e.g., Ferraz-Mello 1988; Sessin \& 
Bressane 1988; Tsuchida 1990). The leftmost panel recovers the $40^\circ\mbox{--}50^\circ$ 
libration of (279) Thule, determined previously by Tsuchida (1990;
Fig.~3). The other two smaller asteroids show librations with comparable
amplitudes. The last object, (52007) 
2002~EQ47, appears to reside on an unstable orbit outside the J4/3
resonance. Our search thus lead to the detection of two new asteroids in 
this resonance, increasing its population by a factor of three.%
\footnote{While our initial search used {\tt AstOrb} catalogue from
 September~2007, we repeated it using the catalogue as of June~2008.
 No additional J4/3 objects were found.}

\begin{figure}
\begin{center}
 \includegraphics[width=84mm]{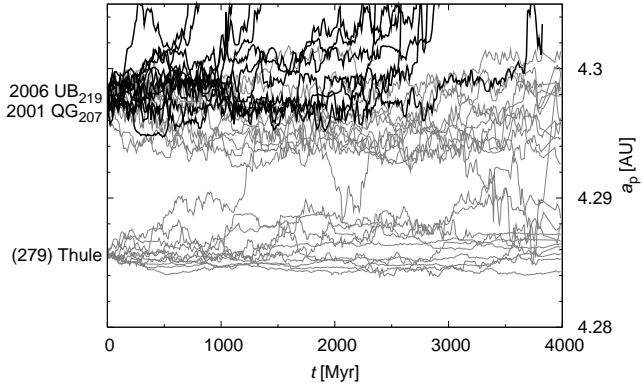}
\end{center}
 \caption{Pseudo-proper semimajor axis $a_{\rm p}$ vs time~$t$
  plot for the 3 orbits of asteroids located inside the J4/3 resonance
  and 10 close clones orbits (placed randomly within the uncertainty ellipsoids)
  for each of them. Thin lines denote the stable
  orbits and thick lines unstable, which escaped from the J4/3 before
  the completion of the simulation at 4\,Gyr.}
 \label{r43-lft_at}
\end{figure}

Results of a long-term numerical integration of the nominal orbits plus
10~close clones, placed within an orbital uncertainty, reveal that the
orbit of (279)~Thule is stable over 4\,Gyr, but the orbits of (186024)
2001~QG$_{207}$ and (185290) 2006~UB$_{219}$ are partially unstable. 
They are not `short-lived' but 45\,\% and 60\,\% of clones, respectively,
escaped before 4\,Gyr. Figure~\ref{r43-lft_at} shows pseudo-proper semimajor
axis vs time for nominal orbits and their clones of all J4/3 objects;
the escaping orbits leave the figure before the simulation was ended at
4\,Gyr. We suspect similar non-conservative effects as mentioned
above for the 20\% fraction of long-term-unstable Hildas to bring these
two Thule members onto their marginally stable orbits.

We would like to point out that it took more than a
century from the discovery of (279) Thule (Palisa 1888; Krueger 1889)
until further objects in this resonance were finally discovered.
This is because there is an anomalously large gap in size of these
bodies: (279) Thule is $127$~km in size with $p_V=0.04$~AU (e.g.,
Tedesco et~al. 2002), while the estimated sizes of (186024) 2001~QG$_{207}$ and
(185290) 2006~UB$_{219}$ for the same value of albedo are $8.9$\,km and $11.8$\,km
only. It will be interesting to learn as much as possible about the 
Thule population in the $H=13\mbox{--}15$\,mag absolute magnitude range using
future-generation survey projects such as Pan-STARRS (e.g., Jedicke
et~al. 2007). Such a completed population may present an interesting
constraint on the planetesimal size distribution 4\,Gyr ago.


\section{Collisional families among Hilda asteroids}\label{fams}

Collisions and subsequent fragmentations are ubiquitous processes since
planets formed in the Solar system. Because the characteristic
dispersal velocities of the ejecta (as a rule of thumb equal to
the escape velocity of the parent body) are usually smaller than the
orbital velocity, the resulting fragments initially reside on nearby
orbits. If the orbital chaoticity is not prohibitively large in the
formation zone, we can recognise the outcome of such past fragmentations
as distinct clusters in the space of sufficiently stable orbital elements.
More than 30 collisional families are known and studied in the main
asteroid belt (e.g., Zappal\`a et~al. 2002) with important additions
in the recent years (e.g., Nesvorn\'y et~al. 2002; Nesvorn\'y,
Vokrouhlick\'y \& Bottke 2006). Similarly, collisional families have
been found among the Trojan clouds of Jupiter (e.g., Milani 1993; 
Beaug\'e \& Roig 2001; Roig et~al. 2008), irregular satellites of 
Jupiter (e.g., 
Nesvorn\'y et~al. 2003, 2004) and even trans-Neptunian objects (e.g.,
Brown et~al. 2007). Mean motion resonances, other than the Trojan
librators of Jupiter, are typically too chaotic to hold stable 
asteroid populations, or the populations were too small to enable
search for families. The only remaining candidate populations are
those in the Jovian first order resonances, with Hilda asteroids
the most promising group. However, low expectations for an existence
of collisional families likely de-motivated systematic
search. Note, the estimated intrinsic collisional probability of
Hilda asteroids is about a factor 3 smaller than in the main
asteroid belt (e.g., Dahlgren 1998; Dell'Oro et~al. 2001) and the
population is more than two orders of magnitude smaller.

In spite of the situation outlined above,
Schubart (1982a, 1991) repeatedly noticed groups of Hilda-type 
asteroids with very similar proper elements. For instance, in his
1991 paper he lists 5 members of what we call Schubart family below and
pointed out their nearly identical values of the proper inclination.
Already in his 1982 paper Schubart mentions a similarity of such
clusters to Hirayama families, but later never got back to the
topic to investigate this problem with sufficient amount of data
provided by the growing knowledge about the J3/2 population.%
\footnote{Schubart lists 11 additional asteroids in the group on
 his website {\tt http://www.rzuser.uni-heidelberg.de/{\~{ }}s24/hilda.htm}, but again he does not go into details of their
 putative collisional origin.}
Even a zero order inspection of Fig.~\ref{r32proper}, in particular the
bottom panel, implies the existence of two large clusters among
the J3/2 population. In what follows we pay a closer analysis to both of them.

\begin{figure}
\begin{center}
 \includegraphics[width=64mm]{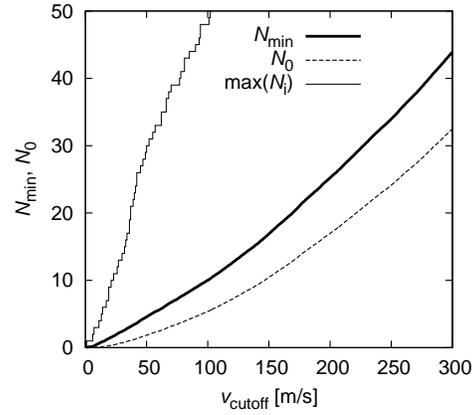}
\end{center}
\caption{The dependence of the minimum number of asteroids $N_{\rm min}$,
  to be considered a statistically significant cluster,
  on the cutoff velocity $v_{\rm cutoff}$ (thick curve).
  The average number $N_0$ and the maximum number ${\rm max}(N_i)$ of
  asteroids, which are closer than $v_{\rm cutoff}$, is shown by dashed and
  thin curves. All quantities are valid for the J3/2 population.
  The fact that ${\rm max}(N_i)$ is much larger than $N_0$ and $N_{\rm min}$
  indicates a presence of a significant cluster (or clusters) among the
  Hilda group.}
 \label{Nmin}
\end{figure}

We adopt an approach similar to the hierarchical-clustering method 
(HCM) frequently used for identification of the asteroid families
in the main belt (e.g., Zappal\`a et~al. 1990, 1994, 2002).
In the first step of our analysis, we compute the number of bodies 
$N_{\rm min}$
which is assumed to constitute a statistically significant cluster
for a given value of the cutoff velocity $v_{\rm cutoff}$. We use a
similar approach to that of Beaug\'e \& Roig (2001): for all asteroids
in the J3/2 resonance we determine the number $N_i(v_{\rm cutoff})$
of asteroids which are closer than $v_{\rm cutoff}$. Then we compute
the average value $N_0 = \bar N_i$. According to Zappal\`a et~al. 
(1994), a cluster may be considered significant if $N > N_{\rm min}
= N_0 + 2\sqrt{N_0}$. The plots $N_0(v_{\rm cutoff})$ and the
corresponding $N_{\rm min}(v_{\rm cutoff})$ for Hilda population
are shown in Fig.~\ref{Nmin}. We use a standard metric
($d_1$ defined by Zappal\`a et~al., 1994), namely:
\begin{equation}
 \delta v = na_p \sqrt{{5\over 4} \left({\delta a_p\over a_p}
  \right)^2 + 2\, (\delta e_p)^2 + 2\, (\delta \sin i_p)^2} \;,
 \label{metric}
\end{equation}
where $(a_p,e_p,\sin i_p)$ are 10~Myr averaged values of the 
resonant pseudo-proper elements (we checked that our results practically
do not depend on the width of this averaging interval).

\begin{figure}
 \includegraphics[width=84mm]{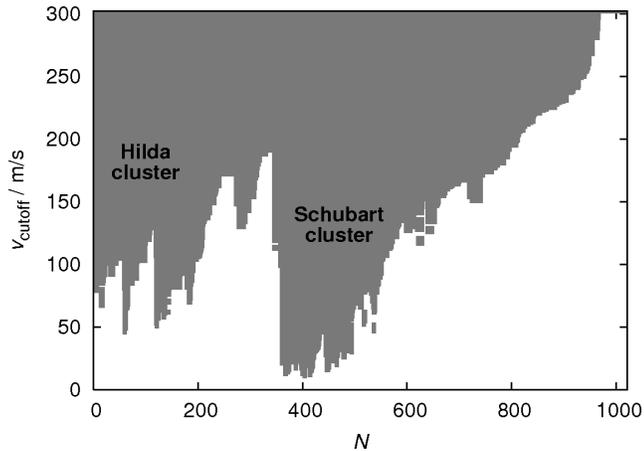}
\caption{A stalactite diagram computed for the J3/2 population (Hildas).
  Two prominent groupings, the Schubart family and the Hilda family,
  are indicated. Every group plotted here has at least 5 or $N_{\rm min}$
  members, whichever is larger (see Fig.~\ref{Nmin}).}
 \label{r32-hcm_stalactite_diagram_min5}
\end{figure}

Next, we construct a stalactite diagram for Hildas in a traditional
way (e.g., Zappal\`a et~al. 1990): we start with (153)~Hilda as the first 
central body and we find
all bodies associated with it at $v_{\rm cutoff} = 300\,{\rm m}/{\rm s}$,
using a hierarchical clustering method (HCM; Zappal\`a et~al. 1990, 1994).
Then we select the asteroid with the lowest number (catalogue designation)
from remaining (not associated) asteroids and repeat the HCM association
again and again, until no asteroids are left. Then we repeat the whole procedure 
recursively for all clusters detected at $v_{\rm cutoff} = 300\,
{\rm m}/{\rm s}$, but now for a lower value, e.g., $v_{\rm cutoff} = 
299\,{\rm m}/{\rm s}$. We may continue until $v_{\rm cutoff} = 0\,
{\rm m}/{\rm s}$, but of course, for too low values of the cutoff velocity,
no clusters can be detected and all asteroids are single. The resulting 
stalactite diagram at Fig.~\ref{r32-hcm_stalactite_diagram_min5}
is simply the asteroid number (designation) vs $v_{\rm cutoff}$ plot: a dot at a 
given place is plotted only if the asteroids belongs to a cluster 
of at least ${\rm max}(5, N_{\rm min}(v_{\rm cutoff}))$ bodies.
We are not interested in clusters with less than 5~members;
they are most probably random flukes.

We can see two prominent clusters among Hildas: the first one around
the asteroid (153)~Hilda itself, and the second one around
(1911)~Schubart. In the remaining part of this Section we discuss each
of them separately.

\begin{figure}
 \includegraphics[width=84mm]{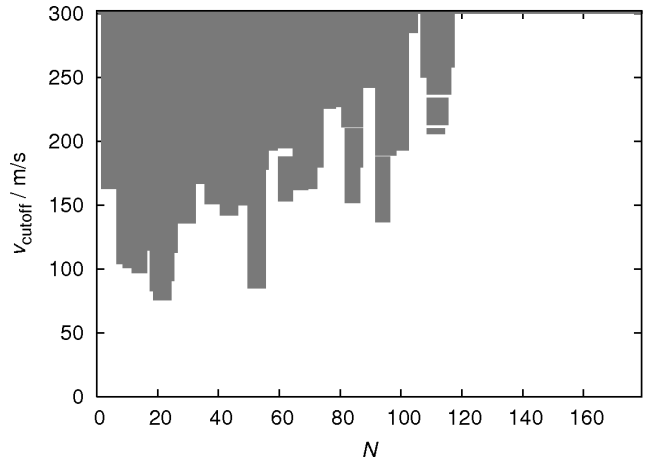}
\caption{A stalactite diagram computed for the long-lived J2/1
  population. There are no prominent groupings; 60~asteroids are not
  associated with any others, even at $v_{\rm cutoff\/} = 300\,{\rm m}/
  {\rm s}$. Every group plotted here has at least 5 members.}
 \label{r21-hcm2007_stalactite_diagram_min5}
\end{figure}

The stalactite diagram constructed in the same way for Zhongguos and Griquas
is shown in Fig.~\ref{r21-hcm2007_stalactite_diagram_min5}. No grouping 
seem to be significant enough to be considered an impact-generated cluster.
This is consistent with the discussion of
the $(a_p,e_p,\sin i_p)$ plots in Section~\ref{hecuba}.

\subsection{Schubart family}\label{schubart_family}

The Schubart family can be distinguished from the remaining population
of Hildas on a large range of cutoff velocities: from 50\,m/s to more than
100\,m/s (Fig.~\ref{r32-hcm_stalactite_diagram_min5}). It merges with the 
Hilda family at 200\,m/s. For the purpose of our analysis we selected
$v_{\rm cutoff} = 60\,{\rm m}/{\rm s}$ as the nominal value. While the 
total number of Schubart family members is not too sensitive to this 
cutoff value, we refrain from using too high $v_{\rm cutoff}$, for which we
would expect and increasing number of interlopers to be associated with
the family, and the family would attain a rather peculiar shape
in the $(a_p, e_p, \sin i_p)$~space.

\begin{figure}
 \includegraphics[width=84mm]{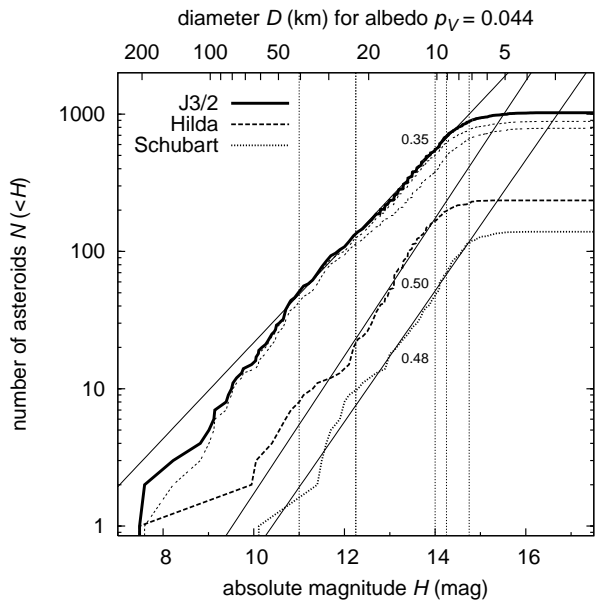}
\caption{Cumulative distributions $N({<}H)$ of absolute magnitudes~$H$
  for the whole J3/2 population (solid) and for the members of the families
  around (153) Hilda (dashed) and (1911) Schubart (dotted). Thin dotted 
  curves denote $N({<}H)$ of the J3/2 population with one and both families
  subtracted. Labels are the best-fitted values of $\gamma$ using 
  $N({<}H)\propto 10^{\gamma H}$ (straight lines) in the interval $(12.25,
  14.25)$ for Hilda and $(12.25, 14.75)$ for Schubart clusters. For sake
  of a rough comparison, the upper abscissa gives an estimate of the sizes
  for an albedo value $p_V=0.044$, average of the J3/2 population.}
 \label{r32_absmag_distribution}
\end{figure}

Figure~\ref{r32_absmag_distribution} shows the cumulative distribution of the
absolute magnitudes for the Schubart family members, compared to the
rest of the J3/2 population. Importantly, the slope $\gamma = (+0.48\pm0.02)$
of the $N({<}H)\propto 10^{\gamma H}$ fit is quite {\em steeper\/} for the Schubart family,
which supports the hypothesis of its collisional origin.

\begin{figure}
 \includegraphics[width=84mm]{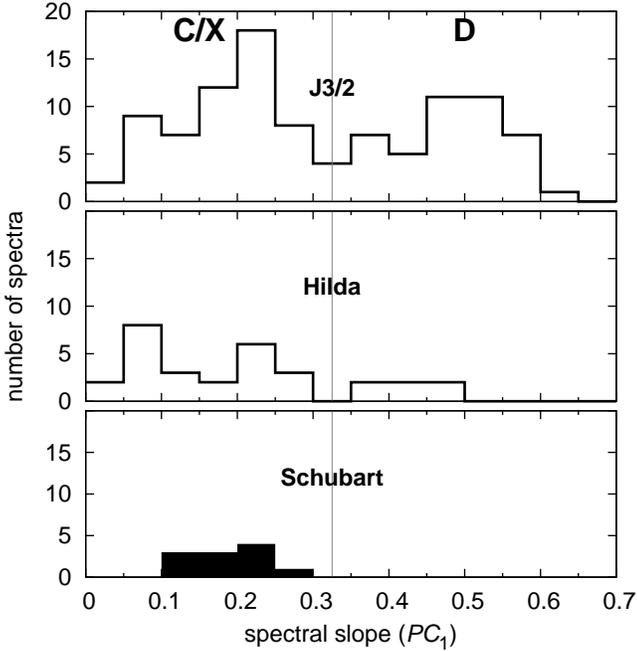}
\caption{Distribution of spectral slopes (PC$_1$ components of the 5~broad-band
  colours) of 153~asteroids in the J3/2 resonance (top), 21~Hilda family members
  (middle) and 4~Schubart family members (bottom). Data computed from the 3rd
  release of the SDSS catalogue of moving objects (ADR3; Ivezi\'c et~al. 2002).
  Note some objects have been observed multiple times by SDSS and the histograms
  show distribution of all observations (rather than averages for a given
  body). The slope values of Hildas range from neutral to steep, with roughly two
  groups separated by the value PC$_1=0.3$: (i) C/X-types with PC$_1<0.3$,
  and (ii) D-types with PC$_1>0.3$. Importantly, the Schubart family members
  are spectrally similar; the median value PC$_1 = 0.20$ corresponds to a C-
  or X-type parent body.}
 \label{r32_wo_schubart.astorb.sloan_pc1_lt0.1_hist}
\end{figure}

We also analysed the available SDSS catalogue of moving objects
(ADR3; Ivezi\'c et~al. 2002). We searched
for the J3/2 asteroids among the entries of this catalogue and computed the
principal component PC$_1$ of the spectrum in the visible band. Note the
PC$_1$ value is an indicator of the spectral slope and allows thus to broadly
distinguish principal spectral classes of asteroids (e.g., Bus et~al. 2002).
Figure~\ref{r32_wo_schubart.astorb.sloan_pc1_lt0.1_hist} shows our results.
The top panel confirms the bimodal character of the J3/2 population
(see also Dahlgren et~al. 1997, 1998, 1999 and Gil-Hutton \& Brunini 2008).
More importantly, though, the bottom panel indicates a spectral homogeneity 
of the Schubart family, placing all members within the C/X taxonomy 
class branch. This finding strongly supports collisional origin of the
Schubart family.

\begin{figure}
\begin{center}
 \includegraphics[width=84mm]{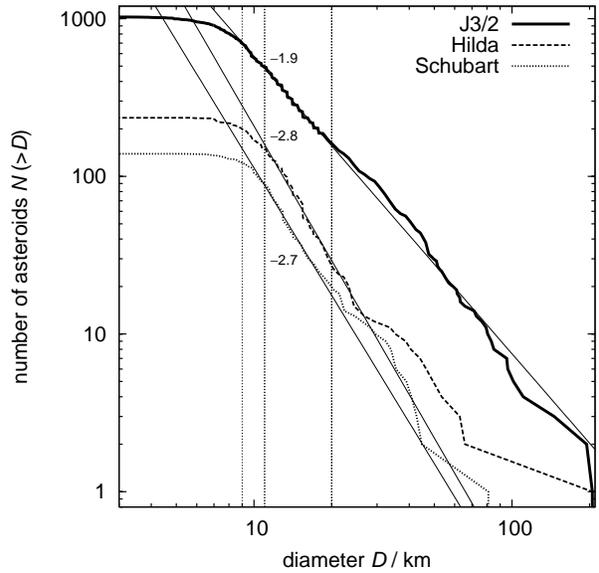}
\end{center}
\caption{Cumulative size distribution $N({>}D)$ for the whole J3/2
  population (solid) and for members of the two suggested collisional
  clusters: Hilda (dashed) and Schubart (dotted). We assumed the
  geometric albedo $p_V = 0.044$ (and $0.025$ in the Schubart-family
  case; see the text) for the conversion of absolute magnitudes $H$ to
  diameters $D$. Labels are the best-fitted values of $\alpha$ using
  $N({>}D)\propto D^\alpha$ (straight lines) in the interval $(11, 20)$\,km 
  for the two families and $(9, 20)$\,km for the J3/2 population.}
 \label{r32_size_distribution}
\end{figure}

Tedesco et~al. (2002) derive $D=80$\,km size for (1911) Schubart,
corresponding to a very low albedo $p_V=0.025$. The same authors
determine $D=38$\,km size of (4230) van den Bergh and exactly the same
albedo; this asteroid is among the five largest in the family.
Assuming the same albedo for all other family members, we can
construct a size-frequency distribution (Fig.~\ref{r32_size_distribution}).
The slope $\alpha\simeq (-2.7\pm0.1)$ fitted to the small end of the distribution,
where we still assume observational completeness, is rather shallow, but
marginally within the limits of population slopes produced in the numerical
simulations of disruptions (e.g., Durda et~al. 2007).%
\footnote{We also mention that so far asteroid disruption simulations
 did not explore cases of weak-strength materials appropriate for the
 suggested C/X spectral taxonomy of the Schubart family parent body.}

If we sum the volumes of the observed members, we end up with a lower 
limit for the parent body size $D_{\rm PB} = 110\,{\rm km}$, provided 
there are no interlopers.
We can also estimate the contribution of small (unobserved) bodies using
the following simple method:
(i)~we sum only the volumes of the observed bodies larger than an assumed
   completeness limit $D_{\rm complete} = 10\,{\rm km}$
   ($V_{\rm complete} = \sum_i {\pi\over 6} D_i^3$);
(ii)~we fit the cumulative size distribution by a power law
   ($\log N({>}D) = \alpha\log [D]_{\rm km} + \beta$;
   $\alpha = -2.68$, $\beta = 4.73$ for the Schubart);
(iii)~we prolong this slope from $D_{\rm complete}$ down to $D_{\rm min} = 0$
   and calculate the total volume of the parent body (provided $\alpha > -3$):
\begin{equation}
 V_{\rm PB} = V_{\rm complete} + {\pi\over 6} 10^\beta {\alpha\over\alpha+3}
  \left[ D_{\rm min}^{a+3} - D_{\rm complete}^{a+3} \right]\; .
\end{equation}
The result is $D_{\rm PB} = \sqrt[3]{{6\over\pi} V_{\rm PB}} \doteq 130\,{\rm km}$,
some sort of an upper limit.
The volumetric ratio between the largest fragment and the parent body is
then $V_{\rm LF}/V_{\rm PB} \doteq 0.2$, a fairly typical value for asteroid
families in the main asteroid belt.
Obviously, the assumption of a single power-law extrapolation of the $N({>}D)$ 
at small sizes is only approximate and can lead to a result with a~$10$\,\% uncertainty. 
However, if we use an entirely different geometric method, developed by Tanga
et~al. (1998), we obtain $D_{\rm PB}\simeq 120\mbox{--}130$\,km,
i.e., comparable to our previous estimate.

What is an approximate size $d_{\rm disrupt}$ of a projectile necessary
to disrupt the parent body of the Schubart family? Using Eq.~(1) from
Bottke et~al. (2005):
\begin{equation}
 d_{\rm disrupt} = \left(2 Q_D^* / V_{\rm imp}^2\right)^{1\over 3}
  D_{\rm target} \; . \label{d_disrupt}
\end{equation}
and substituting $Q_D^* = 10^{5}\,{\rm J}/{\rm kg}$ for the strength
(somewhat lower than that of basaltic objects to accommodate the assumed
C/X spectral type; e.g., Kenyon et~al. 2008 and references therein), 
$V_{\rm imp} = 4.78\,{\rm km}/{\rm s}$ for the typical impact velocity 
(see Dahlgren 1998) and $D_{\rm target} \simeq 130\,{\rm km}$,
we obtain $d_{\rm disrupt} \simeq 25\,{\rm km}$. At this size the projectile 
population is dominated by main belt bodies. Considering also different
intrinsic collisional probabilities between Hilda--Hilda asteroids
($2.3\times 10^{-18}\,{\rm km}^{-2}\,{\rm yr}^{-1}$; Dahlgren 1998)
and Hilda--main belt asteroids ($0.6\times 10^{-18}\,{\rm km}^{-2}\,{\rm yr}^{-1}$),
we find it more likely the Schubart family parent body was hit
by a projectile originating from the main belt.


\subsection{Hilda family}

We repeated the same analysis as in Section~\ref{schubart_family} for the 
Hilda family. The family remains statistically distinct from the
whole J3/2 population in the range of cutoff velocities $(130, 170)\,
{\rm m}/{\rm s}$; we choose $v_{\rm cutoff} = 150\,{\rm m}/{\rm s}$ as 
the nominal value.

The slope $\gamma$ of the cumulative absolute magnitude distribution $N({<}H)$
is $(+0.50\pm0.02)$ (Fig.~\ref{r32_absmag_distribution}), again steeper than
for the total J3/2 population and comparable to that of the Schubart family.
The spectral slopes (PC$_1$) are somewhat spread from flat
(C/X-compatible values; PC$_1< 0.3$) to redder (D-compatible values; PC$_1>0.3$)
--- see Fig.~\ref{r32_wo_schubart.astorb.sloan_pc1_lt0.1_hist}.
Overall, though, the C/X members prevail such that the D-type
objects might be actually interlopers, at least according
to a simple estimate based on the volume of the Hilda family
in the $(a_p, e_p, \sin i_p)$ space, compared to the total volume
of the J3/2 population.


Tedesco et~al. (2002) determine albedos for six family members.
They range from 0.037 to 0.087, but three values are close to the median 
albedo 0.044 of all J3/2 asteroids. We thus consider this value to be
representative of the Hilda family. The corresponding cumulative size distribution
is plotted in Fig.~\ref{r32_size_distribution}. Using the same method as
in Section~\ref{schubart_family} we estimate the size of the parent body
$D_{\rm PB} = 180\mbox{--}190\,{\rm km}$, with $V_{\rm LF}/V_{\rm PB} \simeq 0.8$.
With the model of Tanga et~al. (1999) we would obtain
$D_{\rm PB} \simeq 210$\,km and thus $V_{\rm LF}/V_{\rm PB}\simeq 0.5$.
This family forming event seems to be thus characterized
in between the catastrophic disruption and a huge cratering. The
necessary projectile size is $d_{\rm disrupt} = 50\mbox{--}55$\,km.

While not so prominent as the Schubart family, we consider the group
of asteroids around Hilda a fairly robust case of a collisionally-born 
family too.

\begin{figure}
 \includegraphics[width=84mm]{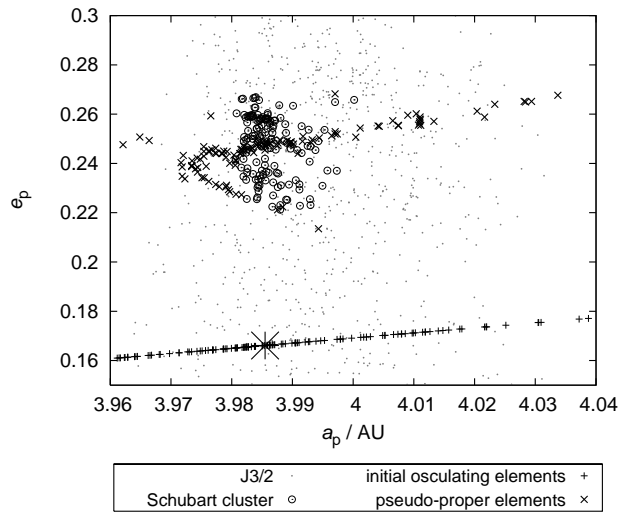}
\caption{The initial osculating elements of an impact-generated swarm
  of 139 fragments at the location of (1911)~Schubart (bottom crosses),
  the corresponding pseudo-proper elements computed from the first My
  of evolution (upper crosses) and the pseudo-proper elements of the
  observed Schubart family (circles). We show here projection onto the
  plane defined by semimajor axis and eccentricity. Dots are the
  pseudo-proper elements of the background J3/2-population asteroids.
  The initial synthetic swarm of asteroids poorly matches the
  observed family: it is both too extent in semimajor axis and too
  compact in eccentricity.}
\label{r32-impact2_ae_inicond}
\end{figure}

\begin{figure}
 \includegraphics[width=84mm]{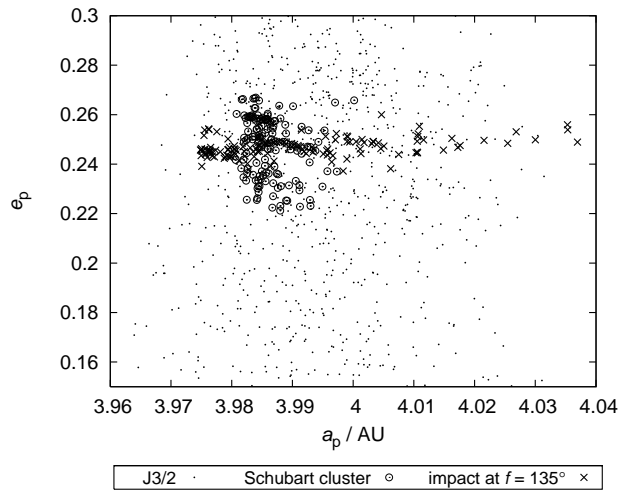}
\caption{The same as Fig.~\ref{r32-impact2_ae_inicond},
  but for a different impact geometry ($f=135^\circ$, $\omega+f =
  180^\circ$). This choice of $f$ maximizes the initial spread
  of the synthetic family in proper eccentricity. In this case we do not
  show the initial osculating orbital elements.}
 \label{r32-impact4_f0_f135_ae}
\end{figure}


\subsection{Simulated disruption events}\label{impacts}

In order to asses some limits for the age of the Schubart and Hilda
families, we perform a number of numerical tests. In particular, we
simulate a disruption of a parent body inside the resonance
and numerically determine the long-term orbital evolution of fragments.
The evolved synthetic family at different time steps is then 
compared with the observed family. Ideally, this approach should
allow to constrain the time elapsed since the family formed.

As a first step, we need to create a synthetic family inside the 
resonance. We use current orbital elements of the largest family
member, (1911) Schubart in this case, as representative to the parent body
and only allow changes in the true anomaly $f$ and in the argument
of pericentre $\omega$ at the break-up event.
By changing these two geometric parameters we can produce different
initial positions of the fragments in the orbital element space.
For sake of our test, fragments are assumed to be dispersed
isotropically with respect to the parent body, with a velocity
distribution given by the model of Farinella et~al. (1993, 1994).
The number of fragments $dN(v)$ launched with relative velocities
in the interval $(v,v+dv)$ is given by
\begin{equation}
 dN(v) = C v (v^2 + v_{\rm esc}^2)^{-(\kappa+1)/2}\, dv\;,
\end{equation}
with $C$ a normalization constant, $v_{\rm esc}$ the escape velocity
{}from the parent body and $\kappa = 3.25$. To prevent excessive
escape velocities we introduce a maximum allowed value $v_{\rm max}$.
Nominally, we set $v_{\rm max}=200$~m/s, but in the next Section~\ref{yarkovsky_e}
we also use restricted values of this parameter to test sensitivity
of our results to initial conditions.

To simulate an impact that might have created the Schubart family,
we generated velocities randomly for 139~fragments with 
$v_{\rm esc} = 65\,{\rm m}/{\rm s}$ (note the number of fragments
in the synthetic family is equal to the number of the Schubart family
members). The resulting swarm of fragments is shown in 
Fig.~\ref{r32-impact2_ae_inicond},
for the impact geometry $f = 0^\circ$ and $\omega+f = 180^\circ$.
We show both the initial osculating orbital elements and the pseudo-proper 
elements.

The synthetic family extends over significantly larger range
of the semimajor axis than the observed Schubart family, but all 
fragments still fall within the J3/2 resonance. The eccentricity 
distribution is, on the other hand, substantially more compact.
Only the distribution of inclinations of the synthetic family
roughly matches that of the observed family. We verified this holds
also for other isotropic-impact geometries (such as $f = 135^\circ$
and $\omega+f = 180^\circ$ shown in Fig.~\ref{r32-impact4_f0_f135_ae}).
The peculiar shape of the synthetic family in the pseudo-proper
element space $(a_p, e_p)$ is an outcome of the isotropic disruption,
simply because some fragments fall to the left from the libration centre of
the J3/2 resonance (at 3.97\,AU) and they are `mapped' to the right.
This is because the pseudo-proper elements are the maxima and minima of
$a$ and $e$, respectively, over their resonant oscillations.

\begin{figure}
 \includegraphics[width=84mm]{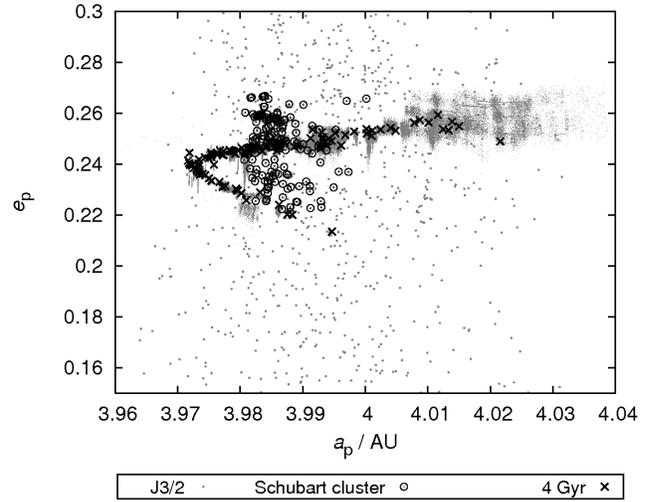}
\caption{The synthetic family from Fig.~\ref{r32-impact2_ae_inicond}
  evolved over 4\,Gyr: the grey dots show evolutionary tracks of the
  fragments in the pseudo-proper orbital element space. Overall, stability
  of the J3/2 resonance makes many fragments to stay very close to their
  initial values. Only ${\sim}10\,\%$ of fragments with the initial
  extremal values of $a_p$ (and thus the libration amplitude) escape
  from the resonance during the simulation. This helps in part to
  reduce the mismatch with the observed family (circles) in semimajor
  axis, but is not sufficient to attain the Schubart-family full
  eccentricity dispersion.}
 \label{r32-impact2_ae_4Gyr}
\end{figure}

The initial configuration of the synthetic family was propagated for
4\,Gyr, using the integrator described in Section~\ref{popul}.
At this stage, we use only the gravitational perturbations from
the 4 exterior giant planets. We performed such simulation for several
impact geometries, as determined by $f$ and $\omega$, with similar results.

Figure~\ref{r32-impact2_ae_4Gyr} shows the long-term evolution of the synthetic family.
Because the family resides mostly in the stable zone of the J3/2 resonance,
only little evolution can be seen for most of the bodies. This is in accord with
findings of Nesvorn\'y \& Ferraz-Mello (1997) who concluded that the
stable region in this resonance shows little or no diffusion over
time scales comparable to the age of the Solar system. Only about
$10$\,\% of orbits that initially started at the outskirts of the stable
zone (with large libration amplitudes) escaped from the resonance during
the 4\,Gyr simulation.

The removal of orbits with large semimajor axis $a_p$ helps in part
to reconcile the mismatch with the distribution of the observed Schubart family.
However, the dispersion in eccentricity $e_p$ does not evolve much
and it still shows large mismatch if compared to the observed family.
Even in the case $f = 135^\circ$ ($\omega+f=180^\circ$;
Fig.~\ref{r32-impact4_f0_f135_ae}), which maximizes the initial
eccentricity dispersion of the synthetic fragments, the final value
at 4\,Gyr is about three times smaller than that of the Schubart
family. Clearly, our model is missing a key element to reproduce
the current orbital configuration of this family.

One possibility to resolve the problem could be to release the
assumption of an isotropic impact and explore anisotropies in the initial velocity
field. This is an obvious suspect in all attempts to reconstruct
orbital configurations of the asteroid families, but we doubt
it might help much in this case. Exceedingly large relative velocities,
compared to the escape velocity of the estimated parent body, would be
required. Recall, the fragments located in the stable region of the
J3/2 resonance would hardly evolve over the age of the Solar system.

A more radical solution is to complement the force model, used for the
long-term propagation, by additional effects. The only viable mechanism
for the size range we are dealing with is the Yarkovsky effect.
This tiny force, due to anisotropic thermal emission, has been proved
to have determining role in understanding fine structures of the
asteroid families in the main belt (e.g., Bottke et~al. 2001;
Vokrouhlick\'y et~al. 2006a,b). In these applications the Yarkovsky effect
produces a steady drift of the semimajor axis, leaving other orbital
elements basically constant.

However, the situation is different for resonant orbits.
The semimajor axis evolution is locked by the strong gravitational
influence of Jupiter. For that reason we first ran simplified
simulations with the Yarkovsky forces --- results of these tests
are briefly described in the Appendix~A. 
We next applied the model containing both gravitational and Yarkovsky
perturbations to the evolution of the synthetic family.
Results of these experiments are described in the next Section~\ref{yarkovsky_e}.


\subsection{Yarkovsky drift in eccentricity}
\label{yarkovsky_e}

We ran our previous simulation of the long-term evolution
of the synthetic family with the Yarkovsky forces included.
Our best guess of thermal parameters for bodies of the C/X type is:
$\rho_{\rm s} = \rho_{\rm b} = 1300\,{\rm kg}/{\rm m}^3$ for the surface
and bulk densities,
$K = 0.01\,{\rm W}/{\rm m}/{\rm K}$ for the surface thermal conductivity,
$C = 680\,{\rm J}/{\rm kg}$ for the heat capacity,
$A = 0.02$ for the Bond albedo and
$\epsilon = 0.95$ for the thermal emissivity parameter.
Rotation periods are bound in the 2--12\,hours range.
Spin axes orientations are assumed isotropic in space.
Finally, we assign sizes to our test particles equal to the estimate of
sizes for Schubart family members, based on their reported 
absolute magnitudes and albedo $p_V=0.025$.
The dependence of the Yarkovsky force on these parameters is
described, e.g., in Bottke et~al. (2002, 2006).
We note the uncertainties of the thermal parameters, assigned
to individual bodies, do not affect our results significantly,
mainly because we simulate a collective evolution of more than
100~bodies; we are not interested in evolution of individual
orbits.

\begin{figure}
 \includegraphics[width=84mm]{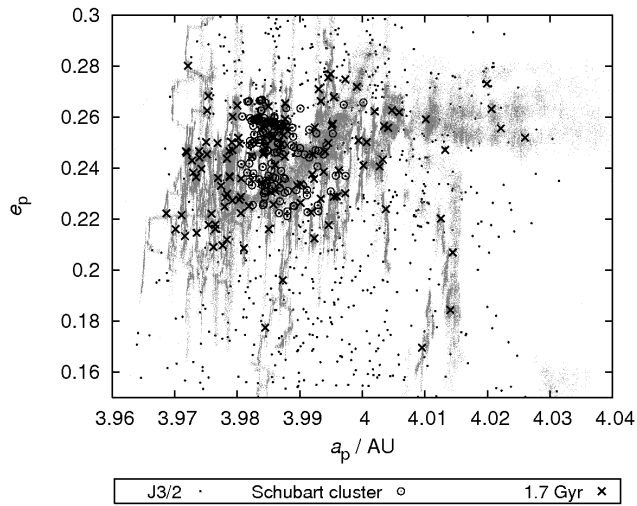}
\caption{The impact-generated swarm from Fig.~\ref{r32-impact2_ae_inicond}
  evolved by planetary perturbations and the Yarkovsky forces, in the
  projection on the pseudo-proper semimajor axis $a_p$ vs eccentricity
  $e_p$ plane. The grey dots indicate the evolutionary tracks over the
  whole 4\,Gyr time span and crosses denote the configuration
  at 1.7\,Gyr, when the eccentricity dispersion of the synthetic family
  particles roughly matches that of the observed Schubart family (circles).}
 \label{r32-impye2_ae_1Gyr}
\end{figure}

We let the synthetic family evolve for 4\,Gyr and recorded
its snapshots every 50~kyr. As discussed in the Appendix~A,
the {\em resonant Yarkovsky effect\/} produces mainly secular changes
of eccentricity. We recall this systematic drift in~$e$
must not be confused with the chaotic diffusion in~$e$. We also
note that inclination of the orbits remains stable, in accord
with a good match of the Schubart family by the initial
inclination distribution.

Because the initial eccentricity dispersion of the synthetic family
is much smaller than that of the observed one, its steady increase
due to the combined effects of the Yarkovsky forces and the resonant lock
gives us a possibility to date the origin of the family
(see Vokrouhlick\'y et~al. 2006a for a similar method applied to families
in the main belt). To proceed in a quantitative way, we use
a 1-dimensional Kolmogorov--Smirnov (KS) test to compare cumulative
distribution of pseudo-proper eccentricity values $e_p$ of the
observed and synthetic families (e.g., Press et~al. 2007).

\begin{figure}
 \includegraphics[width=84mm]{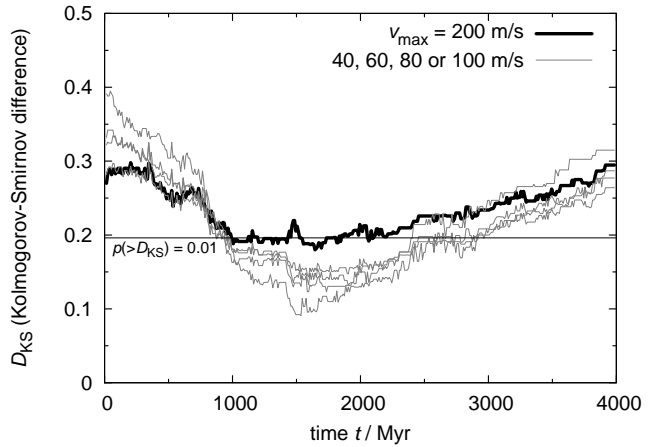}
\caption{The difference between the cumulative distribution
  of eccentricities for the observed Schubart family and our synthetic
  families expressed as the Kolmogorov--Smirnov (KS) distance $D_{\rm KS}$
  vs time $t$. All models assume $f=0^\circ$ and $\omega+f=180^\circ$
  but they have different maximum values $v_{\rm max}$ of
  the ejected particles: $v_{\rm max}=40$, 60, 80, 100 and 200\,m/s
  (labels). For $t<1\,{\rm Gyr}$ and $t>2.4\,{\rm Gyr}$ the KS distance
  is large such that the KS probability is $p({>}D_{\rm KS})<0.01$.
  Thus with a 99\% probability level we can exclude such age for
  the Schubart family. The best matches are achieved for $v_{\rm max}=40$
  and 60\,m/s, for which the most extreme particles have been eliminated
  from the synthetic family.}
 \label{schubart_kstest1_D}
\end{figure}

Figure~\ref{schubart_kstest1_D} shows the KS distance $D_{KS}$
of the two eccentricity distributions as a function of time.
For sake of a test, we also use smaller $v_{\rm max}$ values
of the initial velocity field (essentially, this is like
to start with a more compact synthetic family).
Regardless of the $v_{\rm max}$ value, our model rejects Schubart family ages
smaller than 1\,Gyr and larger than 2.4\,Gyr with a 99\,\% confidence level.
For ages in between 1.5\,Gyr and 1.7\,Gyr the KS-tested likelihood of a similarity
of the synthetic-family and the observed-family $e_p$ distributions
can reach up to $50$\,\%. We thus conclude the most likely age of the
Schubart cluster is $(1.7\pm 0.7)\,{\rm Gyr}$.

\begin{figure}
 \includegraphics[width=84mm]{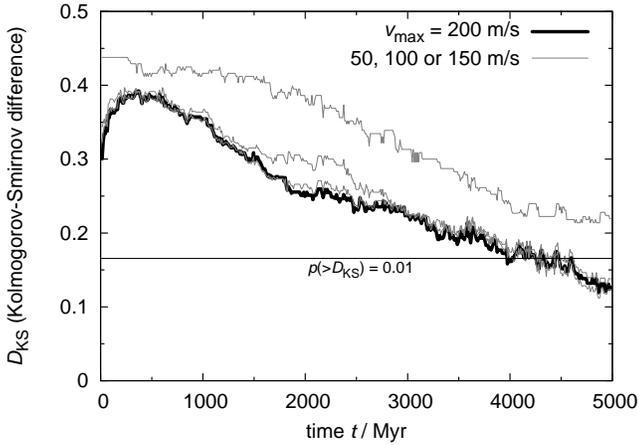}
\caption{The same as Fig.~\ref{schubart_kstest1_D} but for
  the Hilda family. The KS distance~$D_{\rm KS}$ of the observed and modelled
  pseudo-proper eccentricity $e_p$ distribution for several synthetic
  clusters with different maximum velocities $v_{\rm max}$ are plotted
  vs time $t$. In this case, $t\leq 3.5$\,Gyr seems to be ruled out at
  a 99\,\% probability level.}
 \label{hilda_kstest1_D}
\end{figure}

We repeated the same analysis for Hilda family by creating a synthetic
family of 233 particles about (153)~Hilda. In this case we used 
$v_{\rm esc}=110\,{\rm m}/{\rm s}$. The situation is actually very
similar to the Schubart --- there is again a problem with the small
dispersion of eccentricities in case of a purely gravitational model.
Using the model with the Yarkovsky effect, we can eventually fit
the spread of eccentricities and according to the KS test
(Fig.~\ref{hilda_kstest1_D}) the age of the family might be
$\gtrsim 4$\,Gyr. The match is still not perfect, but this
problem might be partly due to numerous interlopers in the family.
We also note that a 10\,\% relative uncertainty of the mean albedo
of the Hilda family members would lead to a 5\,\% uncertainty
of their sizes and, consequently, to a 5\,\% uncertainty of the family age.

The Hilda family seems to be dated back to the Late Heavy Bombardment
era (Morbidelli et~al. 2005; see also Sec.~\ref{migration}).
We would find such solution satisfactory, because the population
of putative projectiles was substantially more numerous than today
(note that a disruption of the Hilda family parent body
is a very unlikely event during the last 3.5\,Gyr).

We finally simulated a putative collision in the J2/1 resonance,
around the asteroid (3789)~Zhongguo (the largest asteroid in the stable
island B). There are two major differences as compared to the J3/2 resonance:
(i) the underlying chaotic diffusion due to the gravitational perturbations
    is larger in the J2/1 resonance (e.g., Nesvorn\'y \& Ferraz-Mello 1997),
    such that an initially compact cluster would fill the whole stable region
    in 1--1.5\,Gyr and consequently becomes unobservable;
(ii)~sizes of the observed asteroids are generally smaller, which together
     with a slightly smaller heliocentric distance, accelerates the Yarkovsky drift
     in $e_p$. The latter effect would likely shorten the time scale to 0.5--1\,Gyr.
Thus the non-existence of any significant orbital clusters in the J2/1 resonance
(Sec.~\ref{fams} and Fig.~\ref{r21-hcm2007_stalactite_diagram_min5})
does not exclude a collisional origin of the long-lived population
by an event older than 1\,Gyr. This would also solve the apparent problem
of the very steep size distribution of the stable J2/1 population (Bro\v{z}
et~al. 2005 and Fig.~\ref{ressfd}). Note the expected collisional
lifetime of the smallest observed J2/1 asteroids is several Gyr
(e.g., Bottke et~al. 2005).


\section{Resonant population stability with respect to planetary
 migration and the Yarkovsky effect}\label{stability}

\begin{figure}
\begin{center}
 \includegraphics[width=64mm]{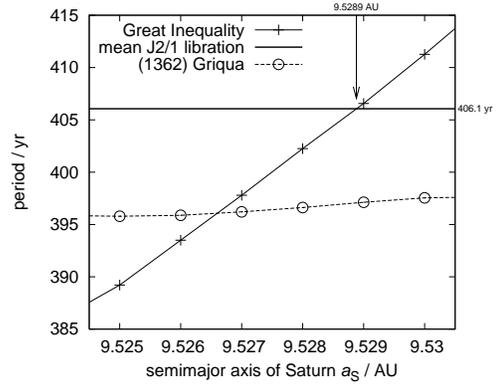}
\end{center}
\caption{Dependence of the period associated with circulation of the
  Great Inequality (GI) angle $5\lambda_{\rm S} - 2\lambda_{\rm J}$
  ($\lambda_{\rm S}$ and $\lambda_{\rm J}$ are mean longitude in orbit
  for Saturn and Jupiter) on the osculating semimajor axis $a_{\rm S}$
  of Saturn (crosses); Jupiter's orbit is fixed.
  Moving Saturn farther away from the location of
  2:5 resonance with Jupiter (at $a_{\rm S} = 9.584\,{\rm AU}$)
  makes the GI period shorter (with today's $a_{\rm S} = 9.5289\,{\rm AU}$
  the GI period is about 880\,yr). In the interval of $a_{\rm S}$ 
  values shown in this figure the GI period becomes comparable to the
  average period $\bar P_{\rm J2/1}$ of librations in the
  J2/1 resonance (solid line). Libration period for a particular case of
  asteroid (1362) Griqua is shown by circles.}
 \label{shifted_Saturn_period_GI}
\end{figure}

We finally pay a brief attention to the overall stability of asteroid
populations in the first-order resonances with respect to different 
configurations of giant planets. We are focusing on the situations when 
the orbits of Jupiter and Saturn become resonant. This is motivated 
by currently adopted views about final stages of building planetary
orbits architecture, namely planet migration in
a diluted planetesimal disk (e.g., Malhotra 1995; Hahn \& Malhotra
1999; Tsiganis et~al. 2005; Morbidelli et~al. 2005; Gomes et~al.
2005 -- these last three references are usually described as the
Nice model). Morbidelli et~al. (2005) proved that the primordial
Trojan asteroids were destabilized when Jupiter and
Saturn crossed their mutual 1:2 mean motion resonance and,
at the same time, the Trojan region was re-populated by particles of 
the planetesimal disk. Since
the mutual 1:2 resonance of Jupiter and Saturn plays a central
role in the Nice model, and since these two planets had to cross
other (weaker) mutual resonances such as 4:9 and 3:7 before they
acquired today's orbits, one can naturally pose a question about
the stability of primordial populations in the first-order mean motion
resonances with Jupiter. Ferraz-Mello et~al. (1998a,b) demonstrated
that even subtler effects can influence the J2/1
population, namely the resonances between the asteroid libration period 
in the J2/1 resonance and the period of the Great Inequality (GI) terms
in planetary perturbations
(i.e., those associated with Jupiter and Saturn proximity to their
mutual 2:5 mean motion resonance; Fig.~\ref{shifted_Saturn_period_GI}). 
A first glimpse to the stability
of the first-order resonance populations with respect to these effects
is given in Section~\ref{migration}.

In Section~\ref{yarkovsky} we also briefly estimate the change
of dynamical lifetimes for small J2/1 and J3/2 bodies caused by
the Yarkovsky effect.

\subsection{Planetary migration effects}\label{migration}

In what follows we use a simple approach by only moving 
Saturn's orbit into different resonance configurations with Jupiter's 
orbit. We do not let orbits of these planets migrate, but
consider them static. With such a crude approach we can only get
a first hint about a relative role of depletion of the asteroid
populations in the first-order resonances (note in reality the
planets undergo steady, but likely not smooth, migration and
exhibit jumps over different mutual resonant states; e.g., supplementary
materials of Tsiganis et~al. 2005).

\begin{figure}
 \includegraphics[width=84mm]{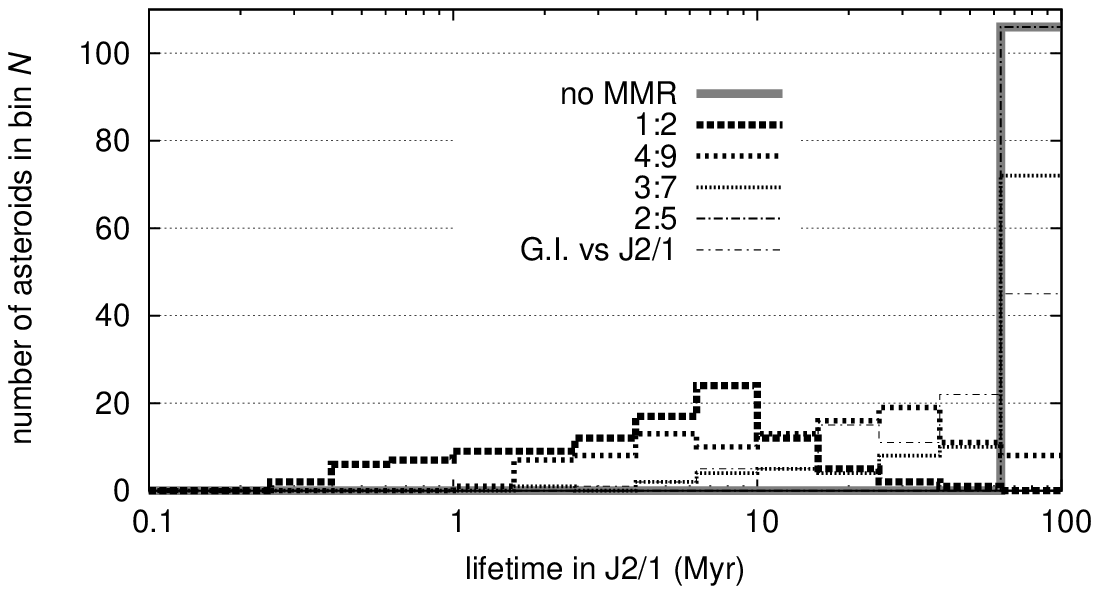}\\
 \includegraphics[width=84mm]{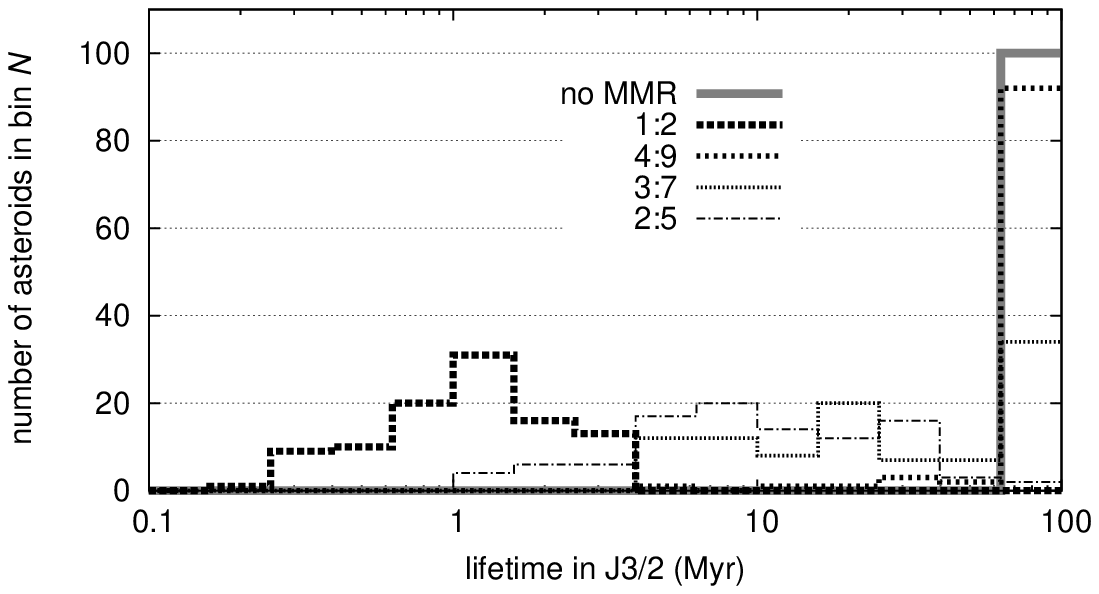}\\
 \includegraphics[width=84mm]{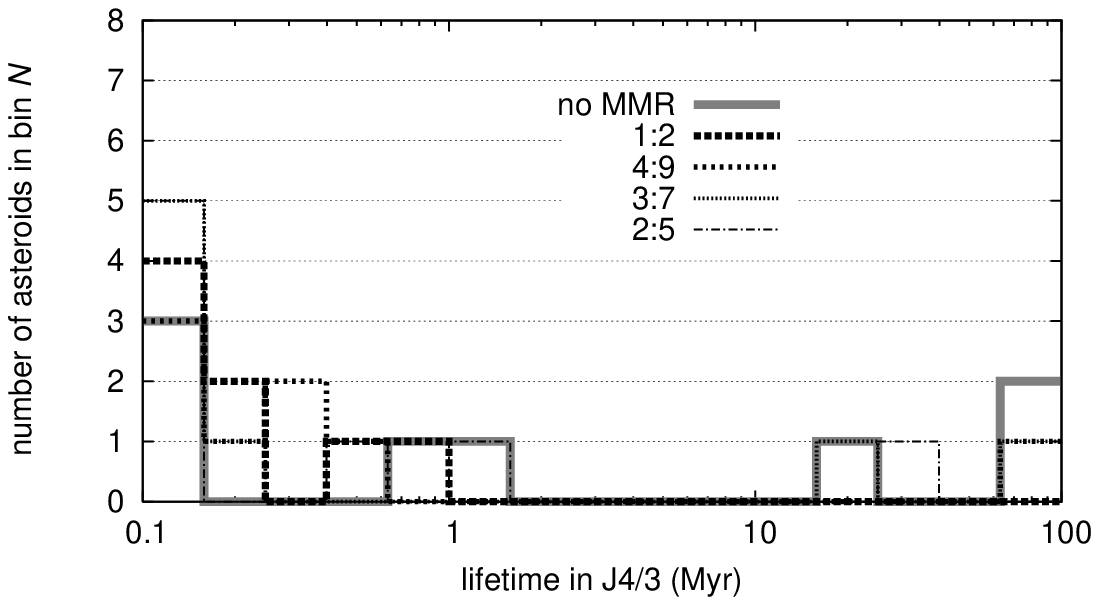}
\caption{Histograms of dynamical lifetimes for asteroids in
  the J2/1 [top], J3/2 [middle] and J4/3 [bottom] resonances,
  in case Jupiter and Saturn are at their current orbits,
  or in a mutual 1:2, 4:9, 3:7, 2:5 resonance,
  or in a Great Inequality resonance (in case of the J2/1 only).
  The histograms were computed for 106~long-lived asteroids
  in the J2/1, first 100~Hildas in the J3/2 and 8 in the J4/3
  (including short-lived).}
 \label{r213243-js_loglftime}
\end{figure}

The results are summarised in Fig.~\ref{r213243-js_loglftime}.
(i)~The Hilda group in the J3/2 resonance is very unstable (on the time scale
$\sim\!1\,{\rm Myr}$) with respect to the 1:2 Jupiter--Saturn resonance;%
\footnote{Jupiter Trojans, which are already known to be strongly unstable
(Morbidelli et~al. 2005), would have the dynamical lifetime
of the order $0.1\,{\rm Myr}$ in this kind of simulation.}
on the contrary J2/1 asteroids may survive several 10\,Myr in this
configuration of Jupiter and Saturn, so this population is not
much affected by this phase by planetary evolution (note, moreover,
that Jupiter and Saturn would likely cross the zone of other mutual
1:2 resonance in $\sim\!1\,{\rm Myr}$ only; e.g., Tsiganis et~al.
2005 and Morbidelli et~al. 2005).
(ii)~The 4:9 resonance has a larger influence on the J2/1
population than on Hildas.
(iii)~In the case of 3:7 resonance it is the opposite:
the J3/2 is more unstable than the J2/1.
(iv)~The Great Inequality resonance does indeed destabilise
the J2/1 on a time scale 50\,Myr.
Provided the last phases of the migration proceed very slowly,
it may cause a significant depletion of the primordial J2/1 population.
In the exact 2:5 resonance, the J2/1 population
would not be affected at all.

\begin{figure}
\begin{center}
 \includegraphics[width=84mm]{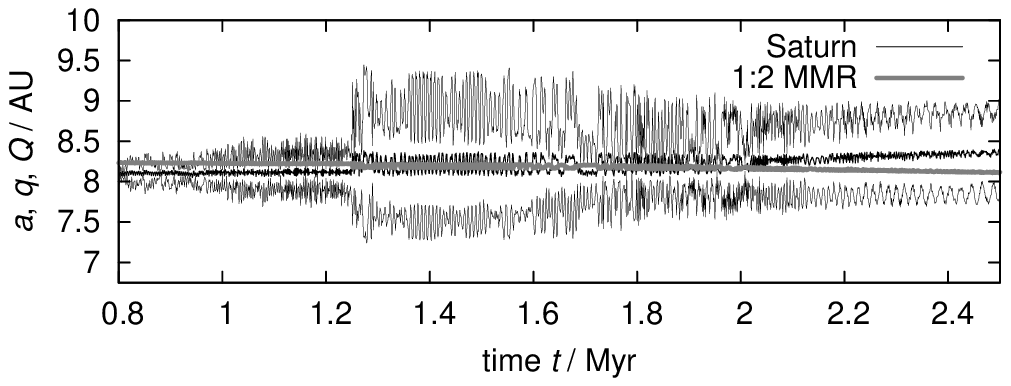}
 \includegraphics[width=84mm]{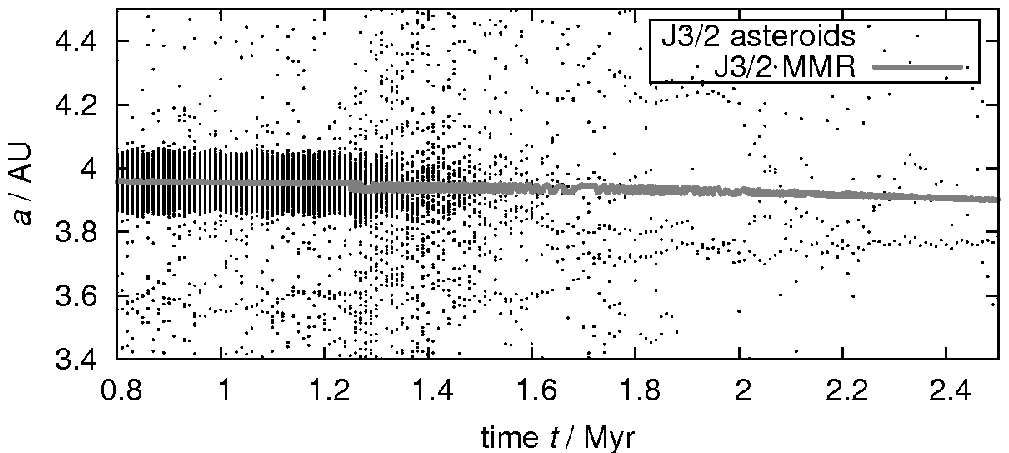}
\end{center}
\caption{An $N$-body simulation of planetary migration driven by dynamically
  cold planetesimal disk beyond Neptune, with $10^3$ particles and total
  mass $50\,M_\oplus$, and including also $10^3$ mass-less particles in the
  J3/2 resonance with Jupiter. Top: The semimajor axis $a_{\rm S}$ of Saturn
  vs time and the position of the 1:2 mean motion resonance with Jupiter
  (estimated from the Kepler law $(1/2)^{-2/3} a_{\rm J}$). Bottom: The same
  for asteroids initially located inside the J3/2 resonance with Jupiter. 
  The J3/2 asteroids are strongly destabilised at the very time of the 1:2
  Jupiter--Saturn resonance crossing ($t = 1.25\,{\rm Myr}$) and none of the
  1000~test particles survived in the J3/2 region after a mere 0.5\,Myr.
  This means more than $99.9\,\%$ depletion of the primordial population.
  None of the planetesimal-disk particles got trapped in the J3/2
  during or after Jupiter's and Saturn's passage through the 1:2 resonance,
  indicating that more particles are needed to study this process.
  We used the Mercury hybrid-scheme integrator (Chambers 1999) for the
  purpose of this test. The gravitational interactions between planets and
  massive planetesimals are accounted for, but planetesimals do not interact
  with each other, nor with mass-less test particles. The time step was
  36~days and the accuracy parameter $10^{-10}$.
  Initial conditions of planets were:
  $a_{\rm J} = 5.2$\,AU,
  $a_{\rm S} = 8.05$\,AU,
  $a_{\rm U} = 12.3$\,AU,
  $a_{\rm N} = 17.5$\,AU,
  with all eccentricities and inclination of the order $10^{-3}$. We took
  the current orbits of Hildas as the initial conditions for our test
  particles.
  Note the destabilisation of the Hilda region is neither sensitive
  to precise initial conditions nor to the mass of the planetesimal disk;
  the only relevant condition is that Jupiter and Saturn cross their mutual
  1:2 resonance.}
\label{j32_pdisc_50ME_cold_I00j_at_planets}
\end{figure}

According to our preliminary tests with a more complete $N$-body model
for planetary migration which includes a disk of $10^3$ planetesimals
beyond Neptune, the strong instability of the J3/2 asteroids indeed
occurs during the Jupiter--Saturn 1:2 resonance crossing 
(see Fig.~\ref{j32_pdisc_50ME_cold_I00j_at_planets}).
A vast reservoir of planetesimals residing beyond the giant planets,
and to some extent also nearby regions of the outer asteroid belt,
are probably capable to re-populate the J3/2 resonance zone at the same time
and form the currently observed populations. This is similar to the
Trojan clouds of Jupiter (Morbidelli et~al. 2005).

\subsection{The Yarkovsky effect}\label{yarkovsky}

In Section~\ref{yarkovsky_e} we already discussed the influence
of the Yarkovsky effect on the families located inside the J3/2
resonance. In course of time it modified eccentricities of their
members, but did not cause a large-scale instability; the families
remained inside the resonance all the time. Here we seek the size
threshold for which the Yarkovsky would case overall instability
by quickly removing the bodies from the resonance.

\begin{figure}
 \includegraphics[width=84mm]{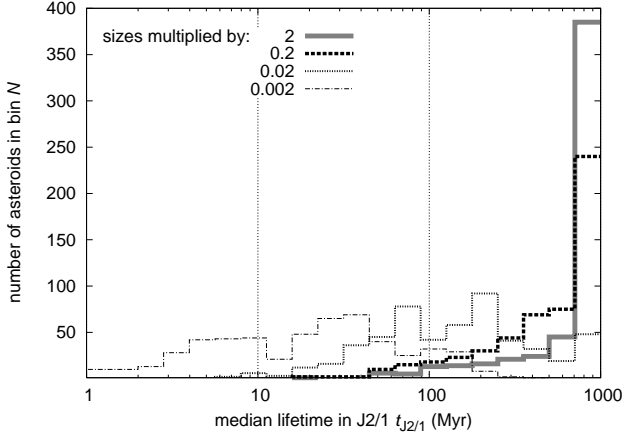}
\caption{Logarithmic histograms of dynamical lifetimes
  for the originally long-lived asteroids in the J2/1 resonance,
  in case the Yarkovsky effect perturbs the orbits.
  The sizes of the objects (4--12\,km) were multiplied by
  2, 0.2, 0.02 and 0.002 for comparison. Note a stronger
  instability starts to occur for the factor 0.02
  (i.e., for the sizes 0.08--0.24\,km). The YORP reorientations
  are not included in this model.}
 \label{r21_zgy_loglftime_radii}
\end{figure}

We perform the following numerical test: we multiply sizes of the 
long-lived J2/1 objects by fudge factors 0.2, 0.02 and 0.002, for 
which the Yarkovsky effect is stronger, and compare respective
dynamical lifetimes with the original long-lived objects. Results
are summarized in Fig.~\ref{r21_zgy_loglftime_radii}. We can conclude
that a significant destabilisation of the J2/1 resonant population
occurs for sizes ${\sim}\,0.1$\,km and smaller (provided the nominal
population has sizes mostly 4--12\,km).

We do not include the YORP effect (i.e., the torque induced by the infrared
thermal emission) at this stage. The YORP is nevertheless theoretically
capable to significantly decelerate (or accelerate) the rotation rate,
especially of the smallest asteroids, which can lead to random reorientations
of the spin axes due to collisions, because the angular momentum
is low in this spin-down state. These reorientations can be simulated
by a Monte-Carlo model with a typical time scale (\v Capek \& 
Vokrouhlick\'y 2004; $R$ is radius in kilometres and $a$ orbital
semimajor axis in AU):
\begin{equation}
 \tau_{\rm YORP} \simeq 25\,{\rm Myr}\, (R/1\,{\rm km})^2\,
  (2.5\,{\rm AU}/a)^2\;. \label{tau_YORP}
\end{equation}
Since the Yarkovsky effect depends on the obliquity value, the systematic
drift would be changed to a random walk for bodies whose spin axis
undergo frequent re-orientations by the YORP effect. We can thus expect
that the YORP effect might significantly prolong dynamical lifetimes 
of resonant objects with sizes ${\sim}\,0.1$\,km or smaller, because
$\tau_{\rm YORP} < 0.25\,{\rm Myr}$ for them.

\begin{figure}
 \includegraphics[width=84mm]{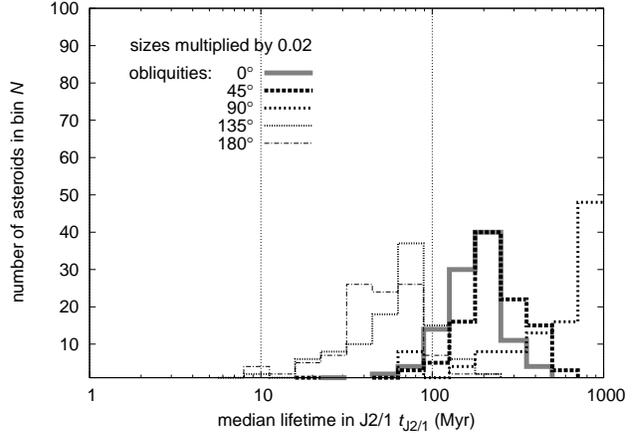}
\caption{The same as Fig.~\ref{r21_zgy_loglftime_radii},
  but now the Yarkovsky effect is calculated only for small bodies
  (0.08--0.24\,km) and varies due to different values of the obliquity
  $\gamma = 0$, $45^\circ$, $90^\circ$, $135^\circ$ and $180^\circ$.
  Note the retrograde bodies (i.e., with a negative drift $da/dt < 0$)
  are significantly more unstable.}
 \label{r21_zgy_loglftime_obliq_0.01}
\end{figure}

We can also check, which orientation of the spin axis makes the
escape from the J2/1 resonance more likely to happen. We consider 
0.08--0.24\,km bodies, 
clone them 5~times and assign them different values of the obliquity
$\gamma = 0^\circ$, $45^\circ$, $90^\circ$, $135^\circ$ and $180^\circ$.
Figure~\ref{r21_zgy_loglftime_obliq_0.01} shows clearly that
the retrograde rotation increases the probability of the escape.
This is consistent with the structure of the J2/1 resonance,
for which low-$a$ separatrix does not continue to $e=0$.
We conclude the remaining very small (yet unobservable)
asteroids inside the J2/1 may exhibit a preferential prograde
rotation.

\begin{figure}
 \includegraphics[width=84mm]{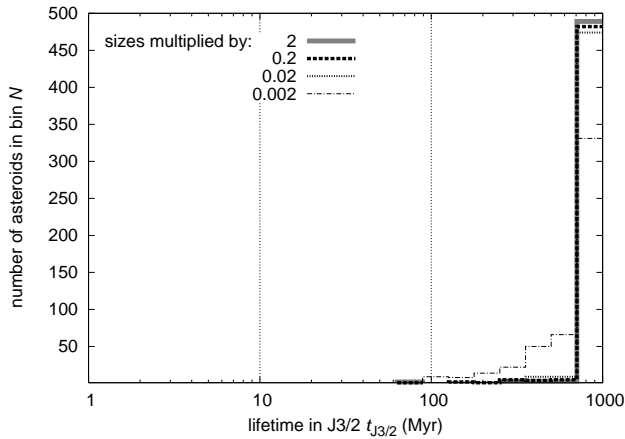}
\caption{The same as Fig.~\ref{r21_zgy_loglftime_radii},
  but for the J3/2 resonance. Instability occurs for the
  sizes multiplied by 0.002 (i.e., 0.02--0.12\,km).
  The YORP effect and corresponding reorientations
  of the spin axes are included in this case.}
 \label{r32_zgy_loglftime_radii}
\end{figure}

We perform a similar simulation for the J3/2 population
(first 100 bodies with sizes 10--60\,km),
but now with YORP reorientations (Eq.~\ref{tau_YORP}) included.
The results (Fig.~\ref{r32_zgy_loglftime_radii}) show
the J3/2 population is much less affected than the J2/1
by the Yarkovsky/YORP perturbation.

We conclude the Yarkovsky/YORP effect may destabilize
the J2/1 and J3/2 bodies only partially on the 100\,Myr time scale
and only for sizes smaller than ${\sim}\,0.1$\,km. It is obviously
a remote goal to verify this conclusion by observations (note the
smallest bodies in these resonances have several kilometers size).
Nevertheless, the dynamical lifetimes of small asteroids determined
in this section, might be useful for collisional models of asteroid
populations, which include also dynamical removal.

\section{Conclusions and future work}

The main results of this paper can be summarised as follows:
(i)~we provided an update of the observed J2/1, J3/2 and J4/3 resonant
populations;
(ii)~we discovered two new objects in the J4/3 resonance;
(iii)~we described two asteroid families located inside the J3/2
resonance (Schubart and Hilda) and provided an evidence that they are of a
collisional origin;
(iv)~we reported a new mechanism how the Yarkovsky effect systematically
changes {\em eccentricities\/} of resonant asteroids; we used this phenomenon
to estimate the ages of the Schubart and Hilda families
($(1.7\pm0.7)$\,Gyr and $\gtrsim 4$\,Gyr respectively);
(v) collisionally-born asteroid clusters in the stable region of J2/1 would
disperse in about 1\,Gyr;
(vi)~20\,\% of Hildas may escape from the J3/2 resonance within 4\,Gyr
in the current configuration of planets;
(vii)~Hildas are strongly unstable when Jupiter and Saturn cross
their mutual 1:2 mean motion resonance.

The J3/2 resonance is a unique `laboratory' --- the chaotic diffusion
is so weak, that families almost do not disperse in eccentricity
and inclination due to this effect over the age of the Solar system.
What is even more important, they almost do not disperse in semimajor
axis, even thought the Yarkovsky effect operates. The drift in~$a$
is transformed to a drift in~$e$, due to a strong gravitational coupling
with Jupiter. We emphasize, this is is not a chaotic diffusion in~$e$,
but a {\em systematic drift in~$e$\/}.

\begin{figure}
\begin{center}
 \includegraphics[width=74mm]{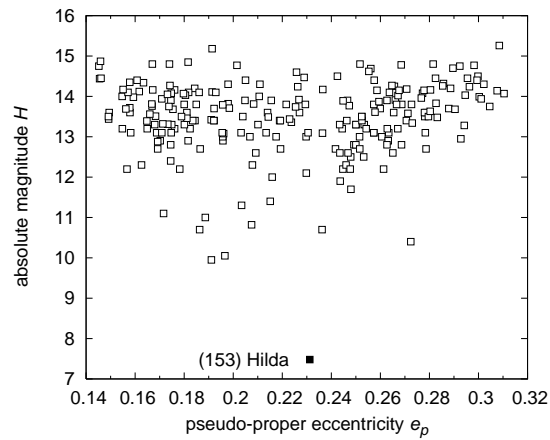}
\end{center}
\caption{Pseudo-proper eccentricity $e_p$ vs absolute magnitude $H$ 
  plot for the members of the Hilda family. The triangular shape and the
  depletion of asteroids in the centre of the family (around $e_p \simeq
  0.22$) is discussed in the text.}
 \label{hilda_eH}
\end{figure}

Another piece of information about the families in J3/2 resonance is
hidden in the eccentricity $e_p$ vs absolute magnitude $H$ plots
(see Fig.~\ref{hilda_eH} for the Hilda family).
The triangular shape (larger eccentricity dispersion of the family 
members for larger $H$) is a well-known combination of two effects:
(i) larger ejection speed, and
(ii) faster dispersal by the Yarkovky forces for smaller fragments.
Interestingly, there is also a noticeable depletion of 
small bodies in the centre of the family and their concentration at 
the outskirts --- a phenomenon known from $(a,H)$ plots of main-belt 
families, which was interpreted as an interplay between the Yarkovsky
and YORP effects (Vokrouhlick\'y et~al.~2006).%
\footnote{The YORP effect tilts the spin axes of asteroids
 preferentially towards obliquity $\gamma = 0^\circ$ or $180^\circ$ and
 this enhances the diurnal Yarkovsky drift due to its $\cos\gamma$
 dependence.}
Indeed the estimated $\sim 4$\,Gyr age for this family matches the time
scale of a YORP cycle for $D\simeq 10$\,km asteroids in the Hilda
region (e.g., Vokrouhlick\'y \& \v{C}apek 2002; \v{C}apek \& 
Vokrouhlick\'y 2004). Vokrouhlick\'y et~al. (2006a) pointed out that
this circumstance makes the uneven distribution of family members most
pronounced.

We postpone the following topics for the future work:
(i)~a more precise age determination for the resonant asteroid families,
based on the Yarkovsky/YORP evolution in the $(e,H)$~space;
(ii)~a more detailed modelling of analytic or $N$-body migration
of planets and its influence on the stability of resonant populations.


\section*{Acknowledgements}
 We thank David Nesvorn\'y, Alessandro Morbidelli and William F. Bottke for
 valuable discussions on the subject and also Rodney Gomes for a constructive review.
 The work has been supported by the Grant Agency of the Czech Republic
 (grants 205/08/0064 and 205/08/P196)
 and the Research Program MSM0021620860 of the Czech Ministry of Education.
 We also acknowledge the usage of the Metacentrum computing facilities
 ({\tt http://meta.cesnet.cz/}) and computers of the Observatory and
 Planetarium in Hradec Kr\'alov\'e ({\tt http://www.astrohk.cz/}).



\appendix

\section[]{Resonant Yarkovsky effect}

\begin{figure*}
\begin{center}
\begin{tabular}{ccc}
 \includegraphics[width=57mm]{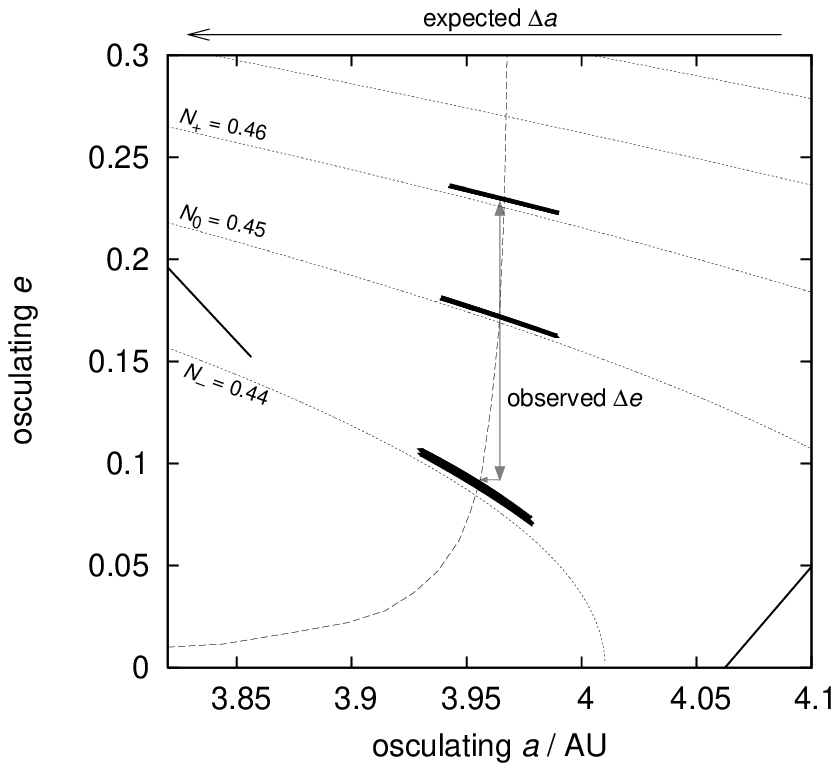} &
 \includegraphics[width=54mm]{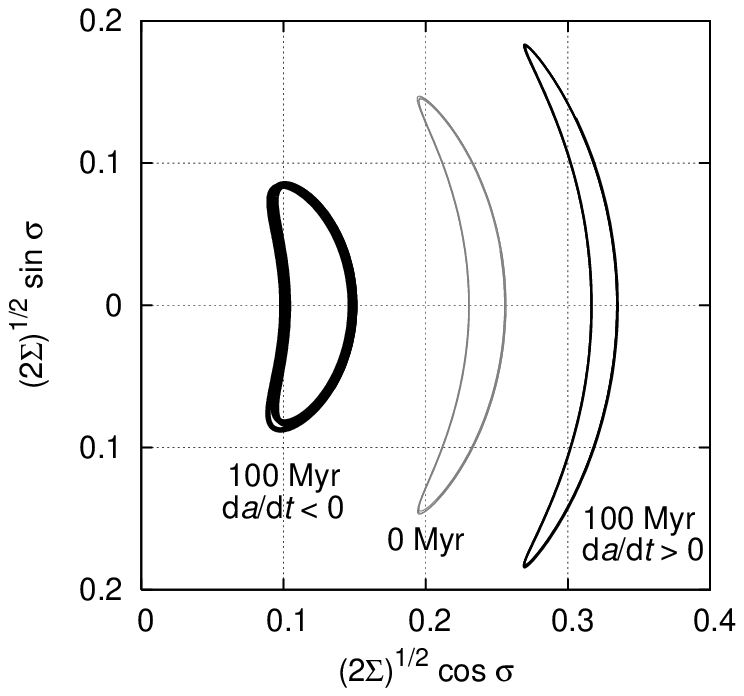} &
 \includegraphics[width=56mm]{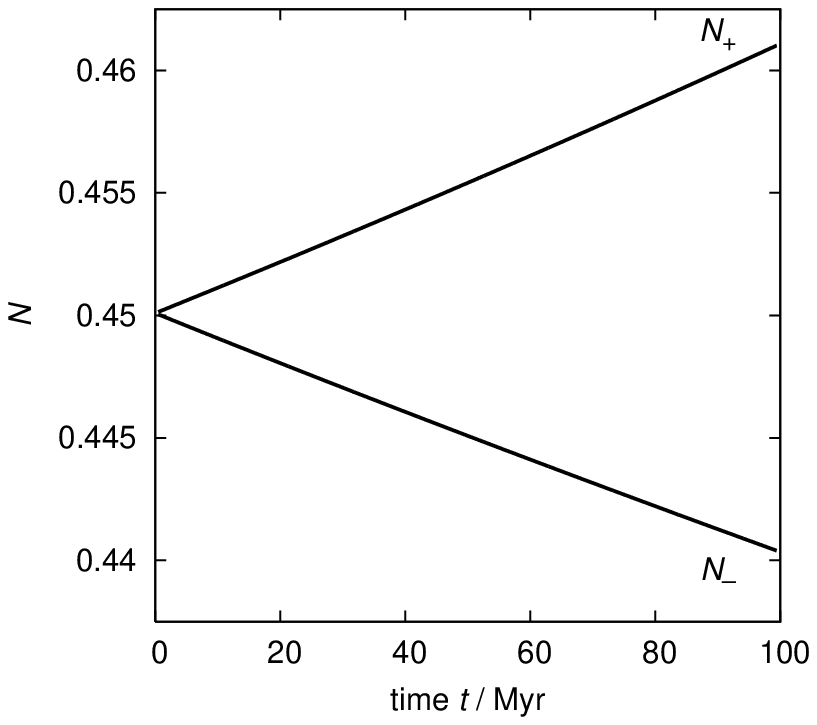}
\end{tabular}
\end{center}
\caption{Orbital evolution of two $D=0.1$~km asteroids in the J3/2 resonance
  within a circular restricted three-body problem (Jupiter on a circular
  orbit). Obliquities $0^\circ$ and $180^\circ$ were assigned
  to the two bodies, such that outside the resonance they would migrate
  by the Yarkovsky forces in opposite direction.
  The expected change $\Delta a$ of the semimajor axis in $100$~My
  is depicted by the arrow on top of the left panel. The left and middle
  panels show $1$\,kyr orbital segments at the beginning and at the end of
  the simulation: (i) in the semimajor axis $a$ vs eccentricity $e$
  projection (left), and (ii) in the projection of Cartesian resonant
  variables $\sqrt{2\Sigma}\, (\cos\sigma,\sin\sigma)$ (see Eqs.~\ref{res11}
  and \ref{res12}; short-period variations have been removed for better
  visibility) (middle). The orbits slowly evolve
  from the initial $N_0\simeq 0.45$ level-curve of the integral given by
  Eq.~(\ref{resint}) to their final values of $\simeq 0.44$ ($N_+$ with
  $da/dt>0$) and $\simeq 0.46$ ($N_-$ with $da/dt<0$),
  respectively (see also the right panel). During this evolution the
  libration centre follows the position of the exact periodic orbit in
  the J3/2 (dashed curve in the left panel). Because the latter has a steep
  progression in $e$ as $a$ changes, orbital evolution is characterised by
  a significant change of the eccentricity $\Delta e$ (also $e_p$) but 
  only a small change in~$a$ (also $a_p$).}
 \label{r32-rtbp_filter_Ssigma_100Myr}
\end{figure*}

The effects of weak dissipative forces, such as the tidal force, gas-drag
force and the Poynting-Robertson force, on both non-resonant and resonant
orbits were extensively studied in the past (e.g., Murray \&
Dermott 1999 and references therein). Interaction of the Yarkovsky drifting
orbits with high-order, weak resonances was also numerically studied
to some extent (e.g., Vokrouhlick\'y \& Bro\v{z} 2002) but no systematic
effort was paid to study Yarkovsky evolving orbits in strong low-order
resonances. Here we do not intend to develop a detailed
theory, rather give a numerical example that can both help to explain
results presented in the main text and motivate a more thorough analytical
theory.

{\em The Yarkovsky effect outside the resonance.\/}
We constructed the following simple numerical experiment:
we took the current orbit of (1911) Schubart as a starting condition
and integrated the motion of two 0.1\,km size objects with extreme
obliquity values $0^\circ$ and $180^\circ$.
Their thermal parameters were the same as in Section~\ref{yarkovsky_e}.
Because the diurnal variant of the Yarkovsky effect dominates the
evolution, the extreme obliquities would mean the two test bodies
would normally (outside any resonances) drift in semimajor axis
in two opposite directions (e.g., Bottke et~al. 2002, 2006).
The two orbits would secularly acquire $\Delta a \simeq +0.25$\,AU or $-0.25$\,AU
in 100~My, about the extent shown by the arrow on top of the left panel
of Fig.~\ref{r32-rtbp_filter_Ssigma_100Myr}.
Since the strength of the Yarkovsky forces is inversely proportional
to the size, we can readily scale the results for larger bodies.

{\em The resonance without the Yarkovsky effect.\/}
If we include gravitational perturbations by Jupiter only, within
a restricted circular three body problem ($e_{\rm J} = 0$), and
remove short-period oscillations by a digital filter, the parameter $N$
from Eq.~(\ref{resint}) would stay constant. The orbit would be
characterized by a stable libration in $(\Sigma,\sigma)$ variables
with about $30^\circ$ amplitude in $\sigma$ (see the curve labelled
0\,Myr in the middle panel of Fig.~\ref{r32-rtbp_filter_Ssigma_100Myr}).

While evolving, some parameters known as the adiabatic invariants are 
approximately conserved (see, e.g., Landau \& Lifschitz 1976; Henrard
1982; Murray \& Dermott 1999). One of the adiabatic invariants is $N$ 
itself. Another, slightly more involved quantity, is the area $J$ enclosed
by the resonant path in the $\sqrt{2\Sigma}\, (\cos\sigma,\sin\sigma)$ space:
\begin{equation}
 J = \oint \sqrt{2\Sigma}\,d\sigma\;. \label{adiab}
\end{equation}
We would thus expect these parameters be constant,
except for strong-enough perturbation or long-enough time scales
(recall the adiabatic invariants are constant to the second order
of the perturbing parameter only).

{\em Resonant Yarkovsky effect.\/}
Introducing the Yarkovsky forces makes the system to evolve slowly.
The lock in the resonance prevents the orbits to steadily drift
away in the semimajor axis and the perturbation by the Yarkovsky forces
acts adiabatically. This is because (i)~the time scale of the resonance 
oscillations is much shorter than the characteristic time scale of the
orbital evolution driven by the Yarkovsky forces, and (ii)~the strength of
the resonant terms in the equations of motion are superior to the strength
of the Yarkovsky accelerations.

We let the two J3/2 orbits evolve over 100~Myr (Fig.~\ref{r32-rtbp_filter_Ssigma_100Myr}).
At the end of our simulation the orbits moved from $N_0\simeq 0.45$ to $N_+\simeq 0.46$
(for the outward migrating orbit) or to $N_-\simeq 0.44$ (for the inward
migrating orbit), respectively. During this evolution, both orbits
remained permanently locked in the J2/1 resonance, librating about
the periodic orbit (dashed line in the left panel of Fig.~\ref{r32-rtbp_filter_Ssigma_100Myr}).
Because the position of this centre has a steep progression in the eccentricity and
only small progression in the semimajor axis, the evolution across
different $N$-planes makes the orbital eccentricity evolve significantly
more than the semimajor axis. This is also seen in the middle panel
of Fig.~\ref{r32-rtbp_filter_Ssigma_100Myr}, where the librating orbits
significantly split farther/closer with respect to origin of coordinates
(note the polar distance from the origin is basically a measure of the eccentricity).
The shape of the librating orbit is modified such that the area $J$
stays approximately constant. We have verified that the relative change
in both adiabatic invariants, acquired during the 100\,Myr of evolution,
is about the same: $\delta N/N \sim \delta J/J \sim 5\times 10^{-2}$.
It is a direct expression of the strength of the perturbation by the
Yarkovsky forces.

We can conclude the Yarkovsky effect results in a significantly different
type of secular evolution for orbits initially inside strong
first-order mean motion resonances with Jupiter. Instead of secularly
pushing the orbital semimajor axis inward or outward from the Sun,
it drives the orbital eccentricity to smaller or larger values,
while leaving the semimajor axis to follow the resonance centre. 

If we were to leave the orbital evolution continue in our simple model,
the inward-migrating orbit would leave the resonance
toward the zone of low-eccentricity apocentric librators.
Such bodies are observed just below the J2/1 resonance.
On the other hand, the outward-migrating orbit would finally increase the
eccentricity to the value when the orbit starts to cross
the Jupiters orbit. Obviously, in a more complete model, with all planets
included, the orbits would first encounter the unstable region
surrounding the stable resonant zone. Such marginally stable
populations exist in both the J3/2 and J2/1 resonances.

\bsp

\label{lastpage}

\end{document}